\documentclass[aps,prl,twocolumn,showpacs,superscriptaddress,groupedaddress]{revtex4}  
\usepackage{graphicx}  
\usepackage{dcolumn}   
\usepackage{bm}        
\usepackage{amssymb}   
\usepackage{overpic}
\usepackage{xcolor}

\begin{document}



\title{Noise-Induced Polarization Switch in Single and Multiplex Complex Networks}

\author{Jan O. Haerter}
\affiliation{Departament de F{\'\i}sica de la Mat\`eria Condensada, Universitat de
  Barcelona, Mart\'{\i} i Franqu\`es 1, 08028 Barcelona, Spain}
  
  \author{Albert D\'{\i}az-Guilera}
\affiliation{Departament de F{\'\i}sica de la Mat\`eria Condensada, Universitat de
  Barcelona, Mart\'{\i} i Franqu\`es 1, 08028 Barcelona, Spain}
    \affiliation{Universitat de Barcelona Institute of Complex Systems (UBICS), Universitat de Barcelona, Barcelona, Spain}
  
  \author{M. \'Angeles Serrano}
  \affiliation{Departament de F{\'\i}sica de la Mat\`eria Condensada, Universitat de
  Barcelona, Mart\'{\i} i Franqu\`es 1, 08028 Barcelona, Spain}
  
  \affiliation{ICREA, Pg.~Llu\'is Companys 23, 08010 Barcelona, Spain}
  
  \affiliation{Universitat de Barcelona Institute of Complex Systems (UBICS), Universitat de Barcelona, Barcelona, Spain}

\date{\today}

\begin{abstract}
The combination of bistability and noise is ubiquitous in complex systems, from biological to social interactions, and has important implications for their functioning and resilience. 
We analyze a simple three-state model for bistability in networks under varying unbiased noise. 
In a fully connected network increasing noise yields a collapse of bistability to an unpolarised state. 
In contrast, in complex networks noise can abruptly switch the polarization state in an irreversible way. 
When two networks are combined through increasing multiplex coupling, one is dominant and progressively imposes its state on the other, offsetting or promoting the ability of noise to switch polarization. 
Our results show that dynamical correlations and asymmetry in dynamical processes in networks are sufficient for allowing unbiased noise to produce abrupt irreversible transitions between extremes, which can be neutralized or enhanced by multiplex coupling.

\end{abstract}

\maketitle

Bistable behavior in systems which can operate in two competing modes is recurrent in a wide range of domains, from cell biology~\cite{ferrell2012bistability,haerter2013collaboration} to social dilemmas~\cite{skyrms2004stag}. 
Concurrently, stochastic fluctuations are inherent to most of these systems. 
In bistable systems of finite size, an obvious effect of noise is to cause fluctuating transitions between the alternate states. 
However, noise may also impact on bistability intrinsically by unfolding a bistable phase into a monostable one. 
This less intuitive effect has been proved in very different systems, like autocatalytic chemical reactions~\cite{togashi2001transitions}
and decision making in financial markets~\cite{kirman1993ants}. 

Even in systems without bistability, noise can induce striking phenomena~\cite{wio2003modern}. 
A paradigmatic example is the Brownian ratchet, where directed motion can be induced by combining Brownian motion with an asymmetric potential~\cite{abbott1999problem}. 
A discretised version of this, the ``Parrondo paradox'', describes the alternation of two processes --- each with losing expectations --- which combine to produce a winning result~\cite{harmer1999parrondo,parrondo2000new,toral2001cooperative}. 
Such noise phenomena are a consequence of temporal correlations in the dynamics. In systems of many constituents, heterogeneity in the patterns of interaction~\cite{ccaugatay2009architecture} and non-trivial topological features, e.g. in complex networks, can further correlate the dynamics 
~\cite{ye2013parrondo,garcia2015regulation,mosquera2015follow,carro2016noisy}. 
For instance, systems of coupled networks can yield monostable-bistable transitions in the mean field Ising model \cite{suchecki2009bistable}. 

Here, we study the interplay between bistability, noise, and the connectivity structure of networks.
We prove that, under quite loose requirements, dynamical node-node correlations subject to unbiased noise can induce abrupt irreversible switching from bistability to monostability. 
The switching disappears when dynamical correlations are removed in the limit of infinite connectivity; noise can then only bring about a neutral mean state in a continuous transition. 
With dynamical correlations, the system remains polarized and the neutral state can only be attained when the dynamics is entirely produced by noise.
Additionally, 
in multiplexes with increasing coupling, there is a dominant layer which attracts the other towards its ``resilient'' extreme. 
We use a simple and general dynamical model for bistability~\cite{dodd2007theoretical,haerter2013collaboration, lovkvist2016dna} under the presence of unbiased noise which affects all constituents. 
We derive an analytic expression for the fixed points of the dynamics for infinite fully-connected systems, and a second-order mean field approximation for regular random graphs, which is solved numerically. 
Scale-free, Watts-Strogatz, and multiplex networks are investigated using numerical simulations.

\begin{figure}[!]
\begin{center}
\begin{overpic}[width=5cm,angle=-90,trim= 0cm 0pt 0pt 0pt,clip]{dummy.pdf}
\put(5,33){\includegraphics[width=5.5cm,trim= 0cm 0cm 0cm 0cm,clip]{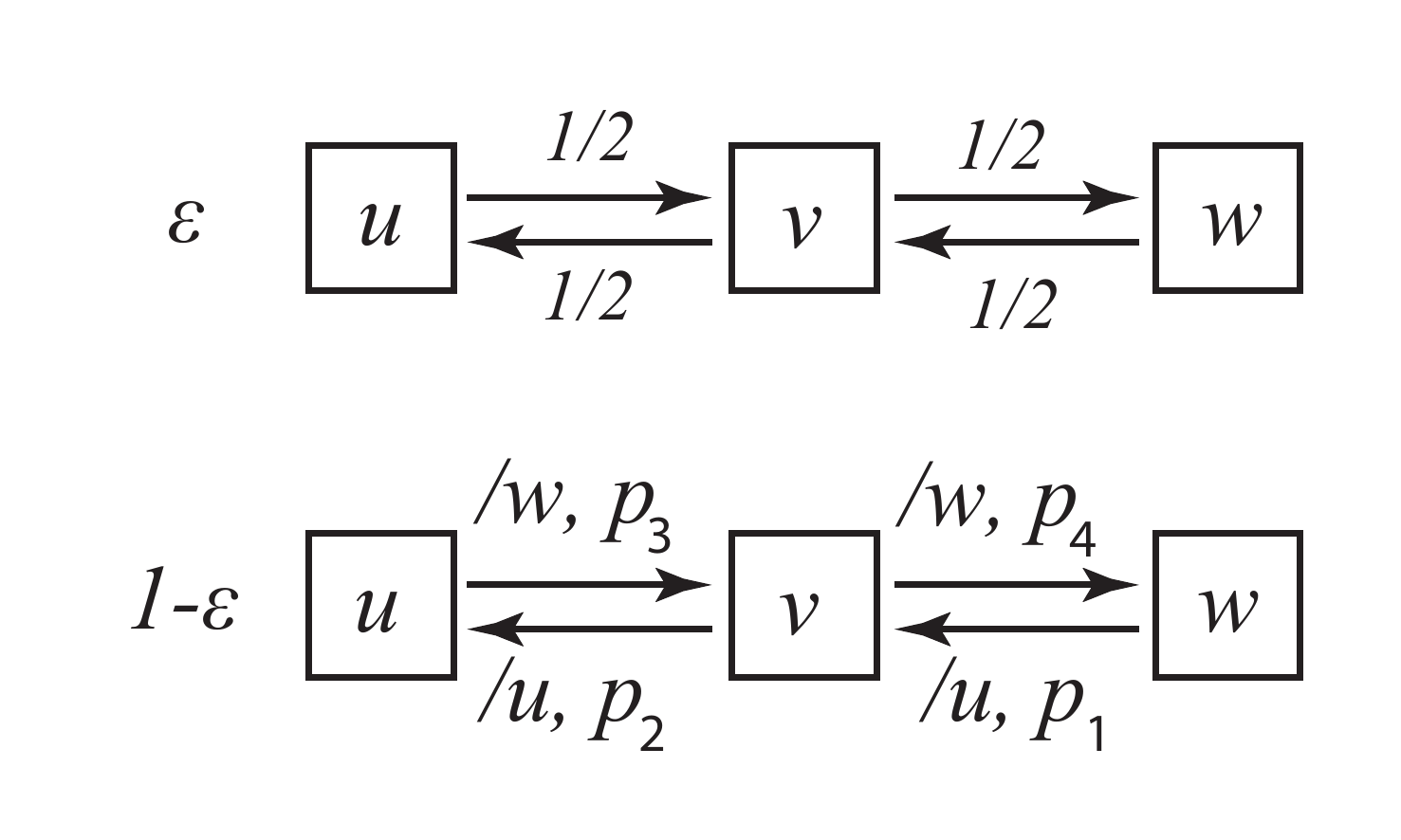}}
\put(-10,70){{\bf a}}
\put(-10,45){{\bf b}}
\end{overpic}
\vspace{-90pt}
\caption{ {\bf Three-state model for bistability.} 
{\bf a.} Unconditional random process. At probability $\varepsilon$, a node is selected and changes its state with probability $1/2$ as shown. {\bf b.} Conditional random process. At probabilty $1-\varepsilon$, transitions are conditional on the state of a selected neighbor, e.g. transition $u\rightarrow v$ for a given node is possible at probability $p_3$ if the selected neighbor is in $w$. No transition takes place otherwise.}
\label{fig:model}
\end{center}
\end{figure}

\noindent
{\bf Model ---} We consider undirected networks of $N$ nodes, where each node can be in one of three states, labeled $u$, $v$ and $w$ (Fig.~\ref{fig:model}) \footnote{Note that we use these three symbols both for the state label and for the state density in the text.}.
These states could encode unmethylated, hemi-methylated, and fully methylated CpG sites in DNA methylation, or left-wing, centrism and right-wing orientation in politics. 
Transitions between the extreme states $u$ and $w$ always visit the intermediate state $v$ and can occur in two ways: 
Type-1 transitions depend only on the state of the affected node (Fig.~\ref{fig:model}a). 
This is analog to Brownian motion, where nodes perform random walks in the space of configurations. 
Type-2 transitions depend on the state of neighbors (Fig.~\ref{fig:model}b). 
The dynamics proceeds as follows: a node $l$ is chosen at random. 
With probability $\varepsilon$ respectively $1-\varepsilon$, a type-1 or type-2 event occurs at $l$. 
In the latter case, one of $l$'s neighbors $m$ is selected at random. 
Depending on the state of $m$, a transition will occur at $l$ with the corresponding probability $p_i$, $i\in \{1,\dots,4\}$ (Fig.~\ref{fig:model}b). 
Note that only four of the nine combinations of states $l$ and $m$ lead to conditional transitions, others are not affected.

\noindent
{\bf Fully connected systems ---}
We first let $p_1$=$p_4$=$1$ and $p_2$=$p_3\equiv p$, leaving only $p$ $\leq$ $1$ free.
We use $\bar{q}$ for the system average of any quantity $q$ as well as the symbol $M\equiv w-u$ to denote net polarization. 
The steady state can be computed exactly for systems consisting of few sites. 
For two nodes connected by a link, $\bar{M}$ maintains its sign for all values of noise and $p$ (Fig.~S1). 
In contrast, for a three-node system forming a triangular loop, a change of average polarization occurs as $\varepsilon$ is increased beyond a certain value $\varepsilon_0$, e.g. $\bar{M}$ $<$ $0$ for $\varepsilon<\varepsilon_0$ and $\bar{M}>0$ for $\varepsilon>\varepsilon_0$. 
We find that $\varepsilon_0$ increases monotonically with $p$ such that even an infinitesimal asymmetry in the model can combine with noise to produce a flip in the polarization sign. 
For larger fully connected systems, this overall change of polarization is maintained. 

However, as $N\rightarrow \infty$, correlations between the states of any two nodes disappear. 
It then becomes appropriate to describe the system's dynamics in a mean field approximation where each node is influenced only by the system averages $\bar{u}$, $\bar{v}$ and $\bar{w}$.
Dropping overbars, the dynamics become $\dot{u}=f(u,w)$ and $\dot{w}=g(u,w)$ \footnote{Here, 
 $f(u,w)=(1-\varepsilon)pu\left(1-2w-u \right)+\varepsilon(1-w-2u)$ and 
 $g(u,w)=(1-\varepsilon)w\left(1-2u-w \right)+\varepsilon(1-2w-u)$.}.
In the steady state ($\dot{u}=\dot{w}=0$), we see that $u=h=w=1/3$ is a fixed point for all $\varepsilon$ and $p$.
Further, for $p=1$ and $\varepsilon<(1+3/\sqrt{p})^{-1}\equiv \varepsilon_c(p)$ the stable fixed point polarization $M\equiv w-u$ becomes $M=\pm\sqrt{1-2\varepsilon'-3\varepsilon'^2}$, where $\varepsilon'\equiv\varepsilon/(1-\varepsilon)$.
For $p<1$, $M$ must be computed numerically and looses its symmetry w.r.t. the origin.
Nevertheless, $\varepsilon_c(p)$ still represents a bifurcation point for the fully-connected system below which two stable polarized fixed points exists. ({\it Details:} Supplement).

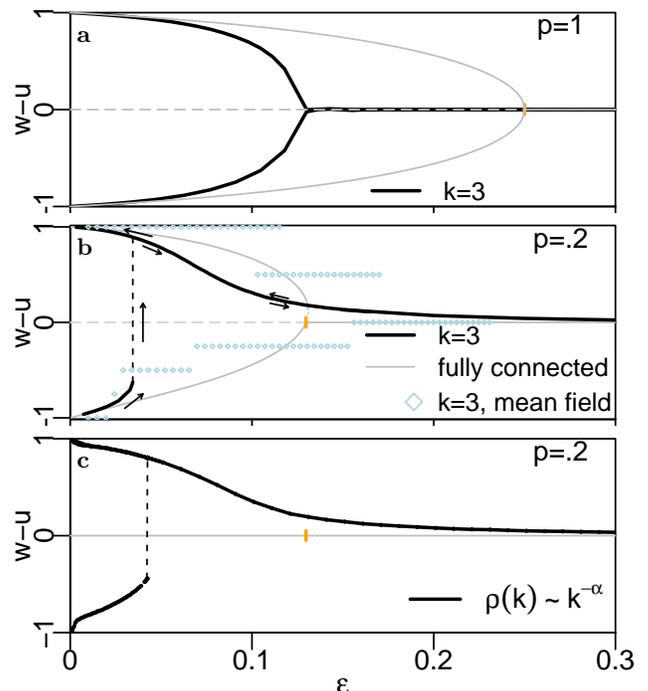
\begin{figure}[t!]
\begin{center}
\begin{overpic}[width=5cm,angle=-90,trim= 0cm 0pt 0pt 0pt,clip]{dummy.pdf}
\put(-22,30){\includegraphics[width=10cm,trim= 0cm 2.5cm 0cm 0cm,clip]{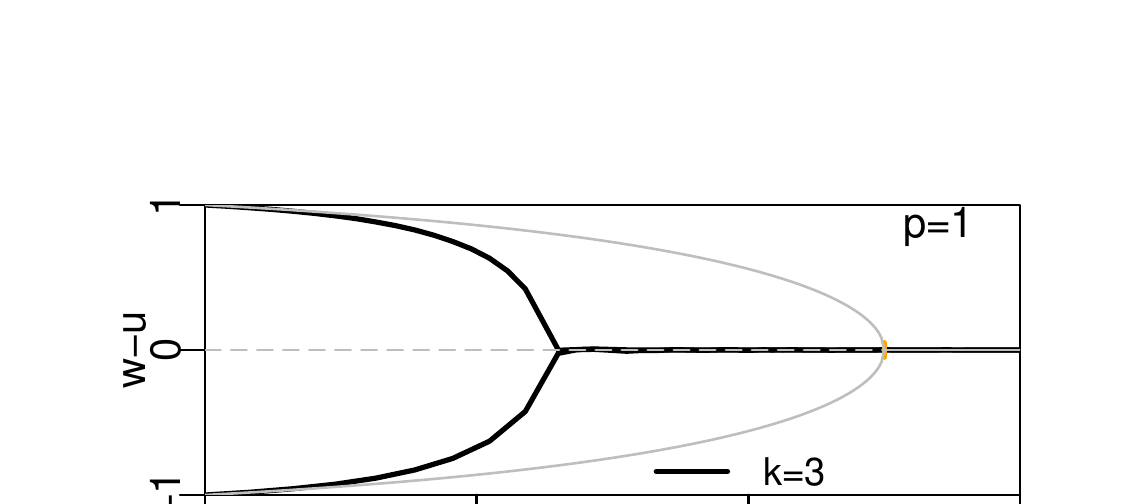}}
\put(-22,-10){\includegraphics[width=10cm,trim= 0cm 2.5cm 0cm 0cm,clip]{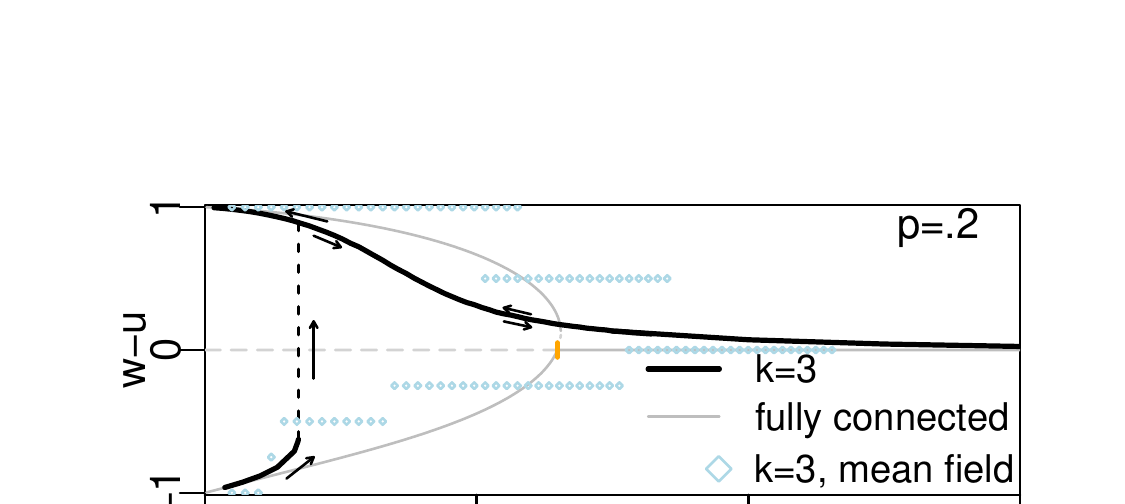}}
\put(-22,-81){\includegraphics[width=10cm,trim= 0cm 0cm 0cm 0cm,clip]{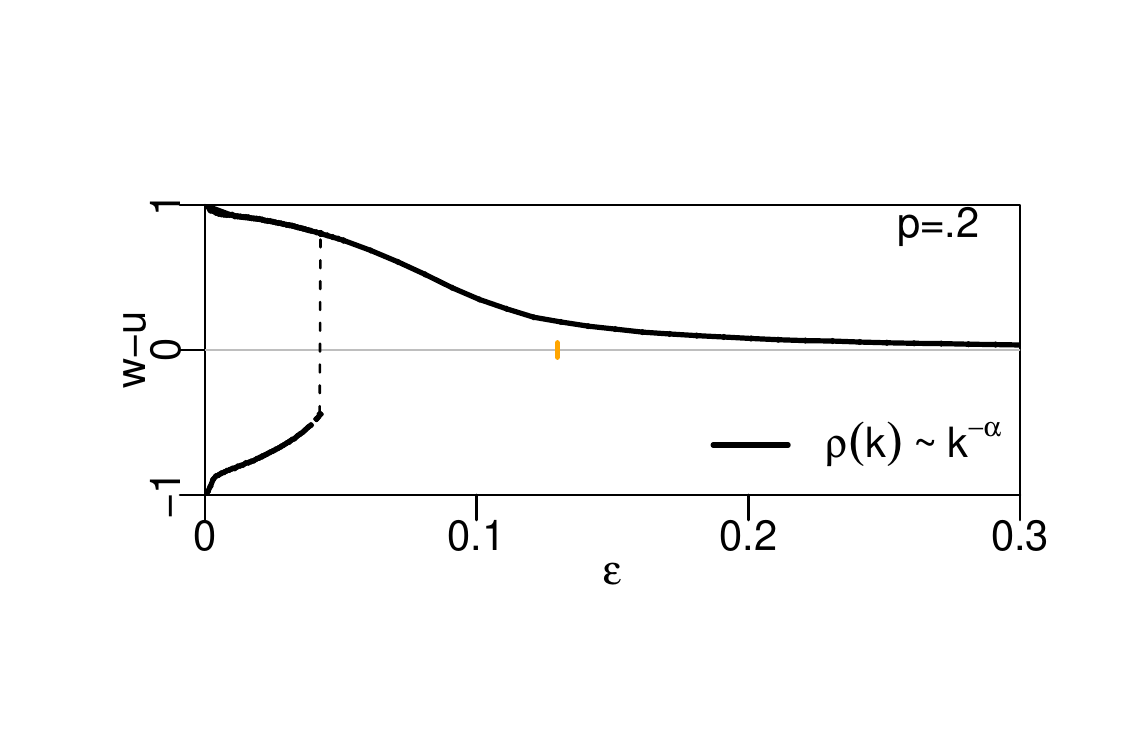}}

\put(5,62){{\bf a}}
\put(5,22){{\bf b}}
\put(5,-18){{\bf c}}

\end{overpic}
\vspace{110pt}
\caption{ {\bf Dynamics in networks.}
{\bf a}, Regular random graph with $k_i=3$ for all nodes $i$ and symmetric system ($p=1$). 
Dependence on $\varepsilon$ for infinite, fully-connected system: solid (dashed) gray lines show stable (unstable) fixed points; 
simulated, sparsely-connected system (bold black lines) when initializing in the two extremal states $\bar{w}=1$ and $\bar{u}=1$, respectively.
Vertical orange mark indicates the bifurcation point $\varepsilon_c(p)=(1+3/\sqrt{p})^{-1}=1/4$ for (i).
{\bf b}, Similar to (a) but for $p=.2$, i.e. for a non-symmetric system.
Dashed vertical line marks the abrupt transition from $\bar{M}<0$ to $\bar{M}>0$.
The vertical orange line marks $\varepsilon_c(.2)\approx .13$.
Light blue symbols mark the distribution maxima obtained through the extended mean field approximation (Fig.~S4) for $k=3$.
Arrows mark the path taken by $\bar{M}$ when starting from $\bar{M}=-1$ and $\varepsilon=0$, then first increasing $\varepsilon$ towards unity and then again reducing $\varepsilon$ to zero.
{\bf c}, Similar to (b) but for scale-free degree distribution with $\alpha=2.4$ and $\langle k \rangle\approx 3$. 
Hysteresis in (c) is similar as for (b), arrows not shown.
System size for simulations: $N=5000$ (see also: Fig.~S8), in (b) near the transition: $N=20000$.
}
\label{fig:correlated_graph}
\end{center}
\end{figure}

\noindent
{\bf Switching through dynamical correlations ---}
How do dynamical correlations influence bistability in the presence of noise? 
To answer this, we explore the range of systems where dynamical correlations are important, hence connectivity is low ($k\ll N$), but system size is large ($N\gg 1$).
We first simulated graphs where each node has identical degree $k$. 
Dynamical correlations are markedly reflected in the phase portrait where trajectories are distorted compared to the fully-connected system, and often cross themselves or one another (Fig. S2). 
For $p=1$, which makes the model symmetric regarding the states $u$ and $w$, we yield a symmetric bifurcation diagram when initializing the system in the extremal states $\bar{M}=\pm 1$, as expected. 
This diagram is qualitatively similar to the one obtained for the fully-connected system (Fig.~\ref{fig:correlated_graph}a). 
However, the bifurcation point in the network, referred to as $\varepsilon^*(p)$, has now moved to considerably lower values of $\varepsilon^*(p=1)\approx .14<\varepsilon_c(p=1)$ (Fig.~\ref{fig:correlated_graph}a) --- an effect of the dynamical correlations retained by demanding small $\langle k\rangle$.
Bistability has become more sensitive to noise. 

Second, we introduce asymmetry ($0\leq p<1$), which means that conditional transitions (i.e. contingent on either $u$ or $w$ at neighboring sites) back-and-forth between $u$ and $v$ are now less frequent than conditional transitions between $v$ and $w$.
This changes the picture qualitatively (Fig.~\ref{fig:correlated_graph}b) as compared to the symmetric case: 
while the system remains bistable at low noise ($\varepsilon<.04$), the lower branch of fixed points is more sensitive to $\varepsilon$ and collapses when $\varepsilon$ is increased. 
The system switches abruptly --- i.e. changes the sign of $M$ --- to a $w$-dominant configuration, and remains polarized there, even as noise is increased further \footnote{To verify that the transition is abrupt, i.e. first order, we simulated the corresponding double well $-ln(\rho(M))$, where $\rho(M)$ is the probability density function of $M$ and checked that the potential well at negative $M$ disappears without merging with that at $M>0$ (Fig.~S13).}. 
Reducing $\varepsilon$ again below $\varepsilon^*(p)$ does not restore the $u$-dominant state, the system remains polarized at $\bar{M}>0$.
Only when $\varepsilon\rightarrow 1$ does the system approach the neutral state where $\bar{M}=0$. 
Henceforth we use the terms ``vulnerable'' and ``resilient'' for the corresponding branches of the system.
In contrast, in fully-connected systems, switching polarization is ruled out for any $0\leq p<1$ (Fig.~S3b). 
$|\bar{M}|$ of a polarized state will decrease as $\varepsilon$ is increased and finally collapse to the neutral state ($\bar{M}=0$) for sufficiently large noise $\varepsilon<1$.

\begin{figure}[t!]
\begin{center}
\begin{overpic}[width=5cm,angle=-90,trim= 0cm 0pt 0pt 0pt,clip]{dummy.pdf}

\put(-8,33){\includegraphics[width=7.0cm,trim= 0cm 2.0cm 0cm 0cm,clip]{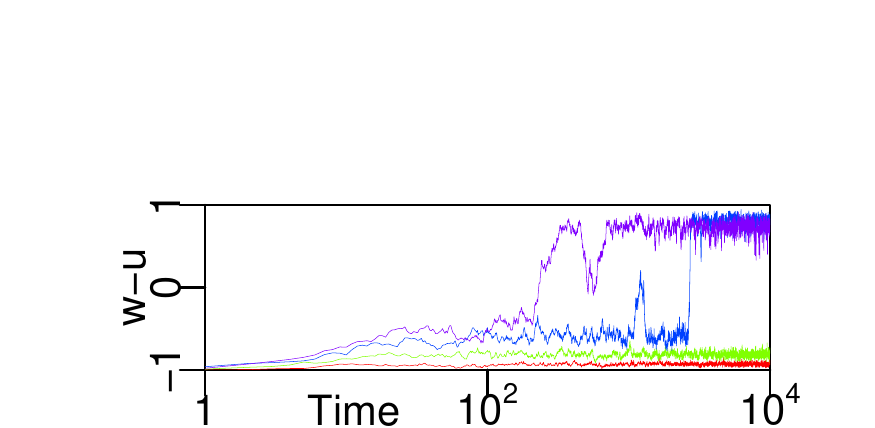}}
\put(-23,25){\includegraphics[width=10.5cm,trim= 0cm 2.5cm 0cm 0cm,clip]{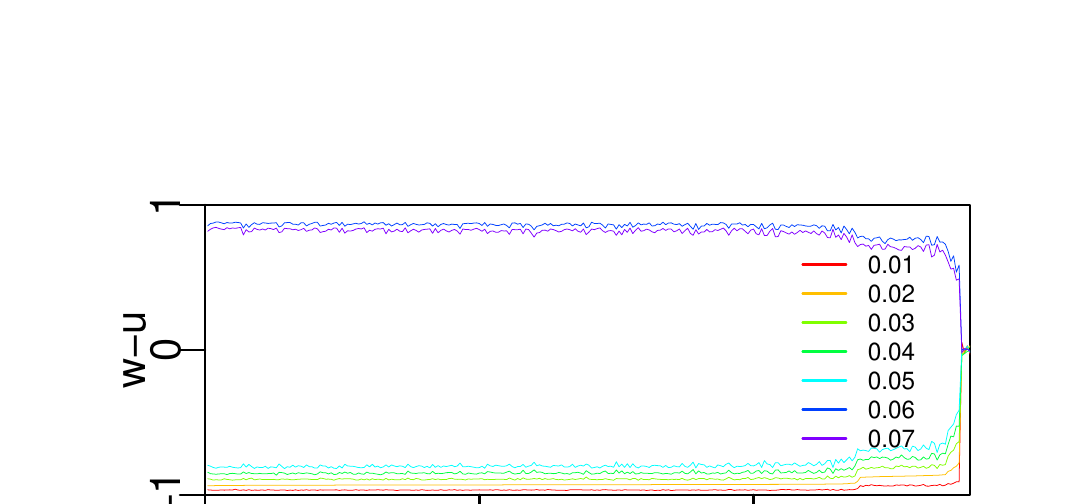}}
\put(-23,-13){\includegraphics[width=10.5cm,trim= 0cm 1.5cm 0cm 0cm,clip]{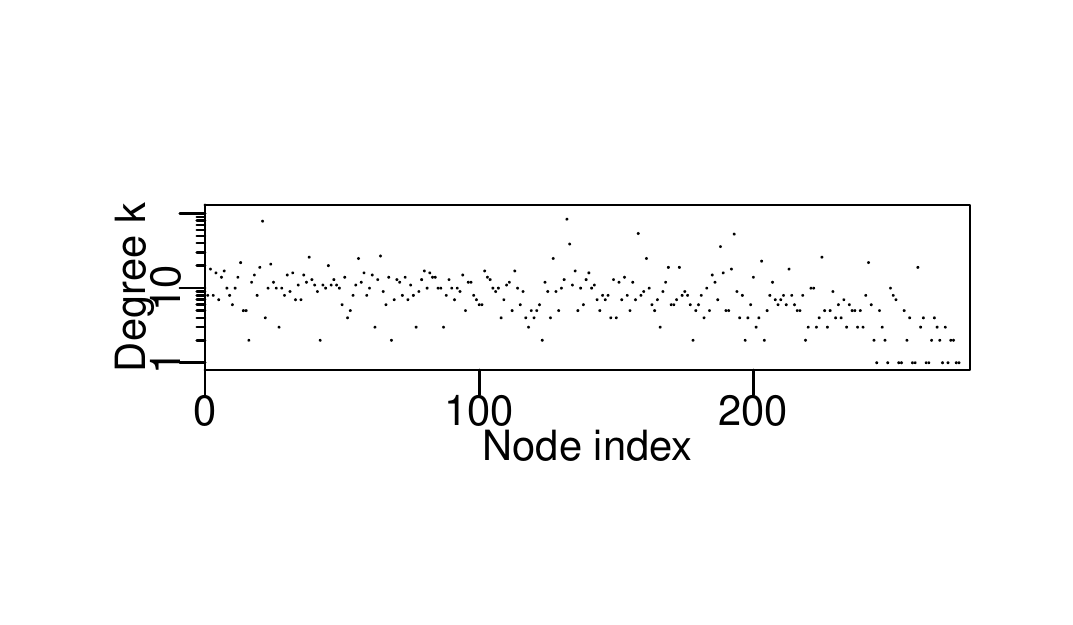}}

\put(6,62){{\bf a}}
\put(6,20){{\bf b}}
\end{overpic}
\vspace{13pt}
\caption{ {\bf Dynamics in the nervous system of \textit{C. elegans}.}  
Electric and chemical synaptic connections \cite{chen2006wiring} (279 nodes).  
{\bf a}, Time-averaged values of $M$ for individual nodes after completion of the transient.
Nodes were sorted by increasing $M$ for the simulation of $\varepsilon=.02$ (orange curve, see legend). 
This ranking was maintained when plotting the results for other values of $\varepsilon$.
Colors (red to purple) correspond to increasing values of $\varepsilon\in \{.01,\dots,.07\}$.
{\it Inset}, Several representative timeseries of network average $\bar{M}\equiv \bar{w}-\bar{u}$ when initializing all nodes at $M=-1$ at $t=0$.
Time is measured in units of system updates.
{\bf b}, Black points indicate the degree of each node. Note the logarithmic vertical axis. 
}
 \label{fig:networks}
\end{center}
\end{figure}

Third, we ask whether such abrupt switches also occur in heterogeneous graphs. 
We synthesized random networks with algebraic degree distributions $P(k)\propto k^{-\alpha}$, $\alpha\approx 2.4$, which gave similar $\langle k\rangle\approx 3$ as in the previous simulation for the regular graph. 
The qualitative features are preserved (Fig.~\ref{fig:correlated_graph}c): bistability is present at low noise but an abrupt switch from $\bar{M}<0$ to $\bar{M}>0$ occurs at a value of noise significantly lower than that of the bifurcation in the fully-connected graph for comparable $p$.
Heterogeneous degree implies that the expected state of any individual node may also be different from that of any other.
Indeed, states not only depend on a node's degree, but additionally on the broader context nodes play within the network (Fig.~\ref{fig:networks}).
When extracting any node's stationary temporal mean state, we find that a node's neighborhood defines its rank, i.e. relative mean state within the network, and this rank is roughly unchanged, even when varying noise or initial conditions. 
Further, random graphs with Poissonian degree distributions as well as the two-dimensional square lattice (Fig.~S6) again gave qualitatively similar behavior as in Fig.~\ref{fig:correlated_graph}b,c.

We also checked the behavior of networks with $\langle k\rangle>3$. 
For fixed $p<1$, we monitored the abrupt transition from bistability to monostability in degree-regular networks of increasing $\langle k\rangle$.
Over approximately two orders of magnitude of $\langle k\rangle$ such transitions are always found (Fig.~S9).
As $\langle k\rangle\rightarrow N$, the transition point $\varepsilon^*(p)$ approaches the bifurcation point $\varepsilon_c(p)$ of the fully-connected system.
We find the difference $|\varepsilon_c(p)-\varepsilon^*(p)|$ to decrease algebraically as $|\varepsilon_c(p)-\varepsilon^*(p)|\sim k^{-s(p)}$ with $0<s(p)<1$ weakly increasing with the asymmetry parameter $p$ (Fig.~S10). 

\noindent 
{\bf Cause of switching ---}
To further explore the origin of the switching we developed a mean-field calculation for sparsely connected networks.
The introductory discussion on the $N=2$, $N=3$, and fully-connected systems highlights that any mean-field approach ignoring dynamical correlations is bound to fail and that it should consider states formed by more than two sites. 
Using a star motif (Fig.~S11) and retaining correlations up to the size of this motif, we write a self-consistent equation for all occupation probabilities of states for the motif. 
The time dependence of occupation probability of each state is then a sum over contributions from internal (within the motif) and external states. 
External states involve conditional probabilities, satisfying compatibility with the internal states.
Solving numerically for the steady state approximates the occupation probabilities $\Pi_i$ ({\it Details}: Supplement).
We find that, for $\langle k\rangle=3$, an asymmetric bifurcation diagram can be obtained when the fixed points are identified as the states of maximum probability (Fig.~\ref{fig:correlated_graph}b).
However, as the degree of the star motif is raised, our mean field approach predicts that one of the branches of fixed points disappears. 

We note the limitation that a mean-field calculation based on a star-motif assumes a tree-like topology and hence disregards effects of clustering and long-range connections. 
To distinguish those effects, we performed additional analysis by implementing Watts-Strogatz networks to interpolate between a regular ring lattice and a random graph. 
As a result of repeated rewiring (Fig.~S7b---d), which gradually reduces clustering by increasing long-range connectivity, the switching $\varepsilon^*(p)$ systematically moves towards larger values of noise (Fig.~S7).
We see this as evidence of the role of long-range connections in enabling the presence of two stable states at small values of noise.  
Bistability can be suppressed by increased locality with associated node-node dynamical correlations. 

This interpretation further implies the existence of dynamical correlations reaching beyond the immediate neighborhood of a given node. 
Using a real-world network, the neural wiring diagram of {\it C. elegans} \cite{chen2006wiring} 
(Fig.~\ref{fig:networks}) and a synthetic scale-free network (Fig.~S14), we consider the long-term average of each individual network node. 
Disregarding the transient phase, we consider only the behavior after $\bar{M}$ settles for either of the polarized states (Fig.~\ref{fig:networks}a). 
We find that a node's degree alone is not a sufficient predictor of its average state within the network (compare Fig.~\ref{fig:networks}a and b). 
Remarkably, even when changing $\varepsilon$, nodes maintain their relative polarization w.r.t. the remaining network, i.e. the curves shown in Fig.~\ref{fig:networks} are approximately still smooth when noise is varied.

\begin{figure}[t!]
\begin{center}
\begin{overpic}[width=5cm,angle=-90,trim= 0cm 0pt 0pt 0pt,clip]{dummy.pdf}
\put(-20,-40){\includegraphics[width=9.5cm,trim= 0cm 0cm 0cm 0cm,clip]{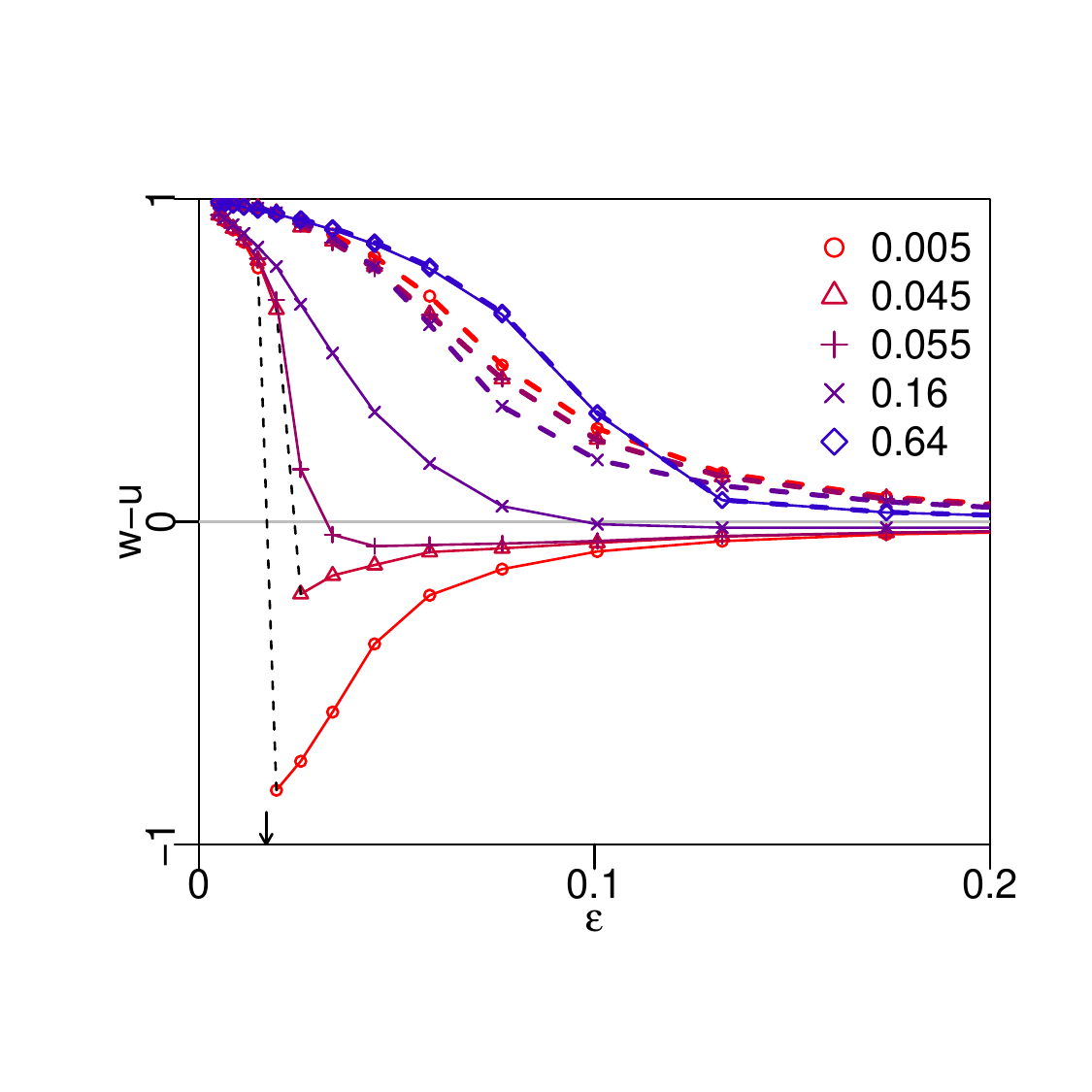}}
\put(27,-22){\includegraphics[width=5.5cm,trim= 0cm 0cm 0cm 0cm,clip]{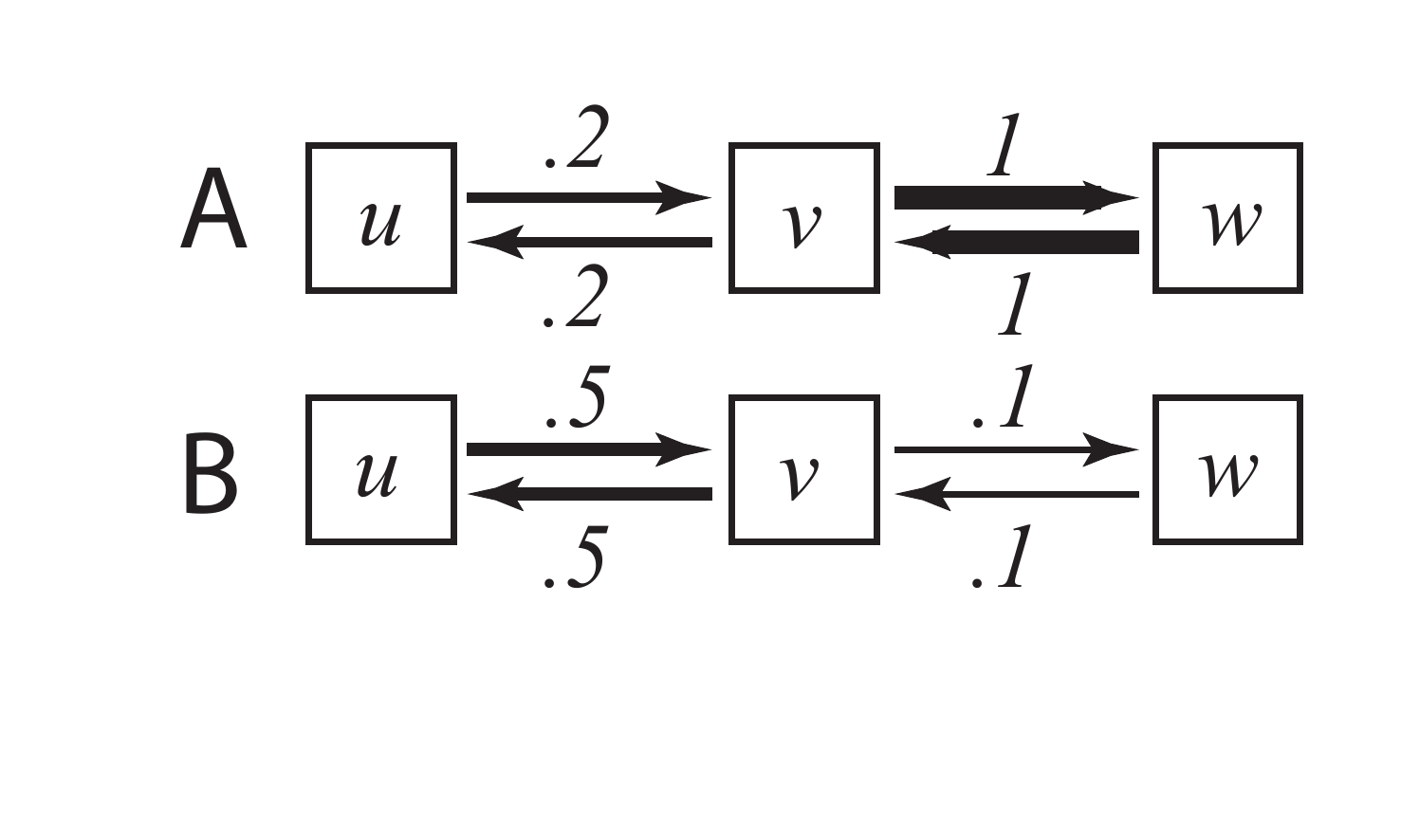}}
\end{overpic}
\vspace{32pt}
\caption{ {\bf Dynamics in a two-layer system with multiplex coupling.}  
Both layers A (dashed lines) and B (solid lines) are initialized at $M=1$ for different values of $\varepsilon$ (see schematic, inset).
The plot legend (top right) indicates various values of coupling $c_f$ increasing from $.005$ to $.64$ for colors ranging from red to blue.
Vertical arrow indicates the approximate position of $\varepsilon^*$ for the isolated system B. $N=2000$.
}
\label{fig:coupling_2layer}
\end{center}
\end{figure}

\noindent
{\bf Two-layer networks ---}
Finally, we combine two network layers A and B, where nodes can change their state by influences from either of the layers.
For simplicity, we consider each layer to consist of statistically equivalent regular random graphs. 
We allow the dynamics in layer A to be defined by the parameters used previously (i.e. $p_1=p_4=1$, $p_2=p_3=.2$) where $\bar{M}>0$ in its resilient branch.
When uncoupled, layer A would hence bahave as described above.
In layer B, rates are generally smaller: $p_1=p_4=.1$, $p_2=p_3=.5$ and swapped regarding $u$ and $w$. 
Hence, B's resilient branch is characterized by $\bar{M}<0$ (Fig.~\ref{fig:coupling_2layer}, inset). 
Both subsystems are initialized as $M=1$ and allowed to reach a steady state. 
We couple A and B by requiring a randomly chosen fraction $c_f$ of nodes to have shared states in the two layers, i.e. their state in one network is copied to the other network at any update. 
The remaining fraction of nodes never transfer their states from one layer to the other. 
At every time step, A or B is chosen at equal probability and a random node of that network is updated as before.

At low noise $\varepsilon$, both systems are $w$-dominant (Fig.~\ref{fig:coupling_2layer}). 
As $\varepsilon$ is increased, the behavior depends on $c_f$. 
For weak $c_f$, B will eventually transition to its resilient configuration, where $M<0$. 
Notably, this transition is still abrupt for small $c_f$ (Fig.~\ref{fig:coupling_2layer}, red curves). 
As $c_f$ is moderately increased (purple curves), the abrupt switch of B is first reduced and eventually disappears but B still shows changing sign of $M$ 
\footnote{The associated reduction of jump size appears to be continuous in the sense that the magnitude of the jump occurring at the zero crossing is gradually reduced as $c_f$ is increased.}.
A, conversely, is also impacted upon by B, since the magnitude of $\bar{M}$ in A is somewhat reduced by the coupling.
As $c_f$ is increased further, even the change of sign finally disappears (blue curves).
For sufficiently large coupling (e.g. $c_f=.64$) the curves of A and B assimilate and the magnitude of $\bar{M}$ at low noise becomes maximal: 
the copying process between layers is now so efficient that they act as a single layer with enhanced effective degree $\tilde{k}>3$ and intermediate effective parameters (compare Fig.~S12). 
Hence, the effect of combining layers is to dilute the correlations between any two neighbors by the larger number of influences from their enlarged neighborhood.
The relative frequencies of updating (e.g. the rates $p_i$) in the two layers determine which of the two will be able to impose its state on the other. 
Variants are explored in Fig.~S15. 
For instance, if network A is initialized in its vulnerable state, it will eventually --- after its own change of polarization --- still dominate the system dynamics by pulling B through the switch.

{\bf Discussion ---} We analyzed noise-induced switching caused by dynamical correlations in a simple three-state model for bistability, but our findings may generalize to other models that induce similar dynamics. 
The switching is a first-order irreversible transition from bistability to monostability. 
Our results suggest that this switching state exists in the intermediate regime where local dynamical correlations are present while nonlocal coupling allows for bistability to be maintained. 
When two systems interact, the dominant one imposes its resilient state on the other as coupling increases. 
This suggests that the propensity of a system to switch polarization under noisy environments can be both counterbalanced or enhanced by its matching with another dominant system. 

Conceptually, our findings could be seen as an extreme case of Parrondo's paradox.
Within this analogy, unbiased noise is a neutral game which transforms a second biased game from overwhelmingly losing to winning when the two are combined. 
In contrast to Parrondo's paradox, the implications of our findings may hence be far less subtle: 
modulating noise could be exploited as a simple mechanism to trigger strong behavioral changes in systems with bistability, e.g. in epigenetics or economics.
For several systems coupled within an overall noisy environment, those with higher rates of conditional reactions will not only be more resilient but may often dominate other systems that are coupled to them. 
Our work might allow for a new perspective on a wide range of transitions in noisy bistable systems such as those that are found in biology or the social sciences.

\section*{Acknowledgments}
J. O. H. thanks Namiko Mitarai and Kim Sneppen for fruitful discussions. 
M. A. S. acknowledges support from the James S. McDonnell Foundation Scholar Award in Complex Systems; the MINECO project no. FIS2013-47282-C2-1-P; and the Generalitat de Catalunya grant no. 2014SGR608. 
A.D.-G. acknowledges financial support from MINECO, Projects FIS2012-38266 and FIS2015-71582 (FEDER), and from Generalitat
de Catalunya Project 2014SGR-608.
All authors acknowledge support from the European Commission FET-Proactive Project MULTIPLEX no. 317532. 

\bibliography{references_parrondo}
\bibliographystyle{apsrev4-1}

\end{document}




\title{Noise-Induced Polarization Switch in Single and Multiplex Complex Networks\\
--- SUPPLEMENTARY MATERIAL ---}

\author{Jan O. Haerter}
\affiliation{Departament de F{\'\i}sica de la Mat\`eria Condensada, Universitat de
  Barcelona, Mart\'{\i} i Franqu\`es 1, 08028 Barcelona, Spain}
  
  \author{Albert D\'{\i}az-Guilera}
\affiliation{Departament de F{\'\i}sica de la Mat\`eria Condensada, Universitat de
  Barcelona, Mart\'{\i} i Franqu\`es 1, 08028 Barcelona, Spain}
    \affiliation{Universitat de Barcelona Institute of Complex Systems (UBICS), Universitat de Barcelona, Barcelona, Spain}
  
  \author{M. \'Angeles Serrano}
  \affiliation{Departament de F{\'\i}sica de la Mat\`eria Condensada, Universitat de
  Barcelona, Mart\'{\i} i Franqu\`es 1, 08028 Barcelona, Spain}
  
  \affiliation{ICREA, Pg.~Llu\'is Companys 23, 08010 Barcelona, Spain}
  
  \affiliation{Universitat de Barcelona Institute of Complex Systems (UBICS), Universitat de Barcelona, Barcelona, Spain}

\pacs{}
\date{\today}

\maketitle

\renewcommand{\thesection}{S\arabic{section}}

\renewcommand{\thefigure}{S\arabic{figure}}
\setcounter{figure}{0}

\renewcommand{\thetable}{S\arabic{table}}
\setcounter{table}{0}

\renewcommand{\theequation}{S\arabic{equation}}
\setcounter{table}{0}

\clearpage

\section{Finite and infinite fully-connected systems}\label{sec:fully_connected}

\begin{figure}[h!]
\begin{center}
\begin{overpic}[width=5cm,angle=-90,trim= 0cm 0pt 0pt 0pt,clip]{dummy.pdf}
\put(-10,0){\includegraphics[width=8cm,trim= 0cm 0cm 0cm 0cm,clip]{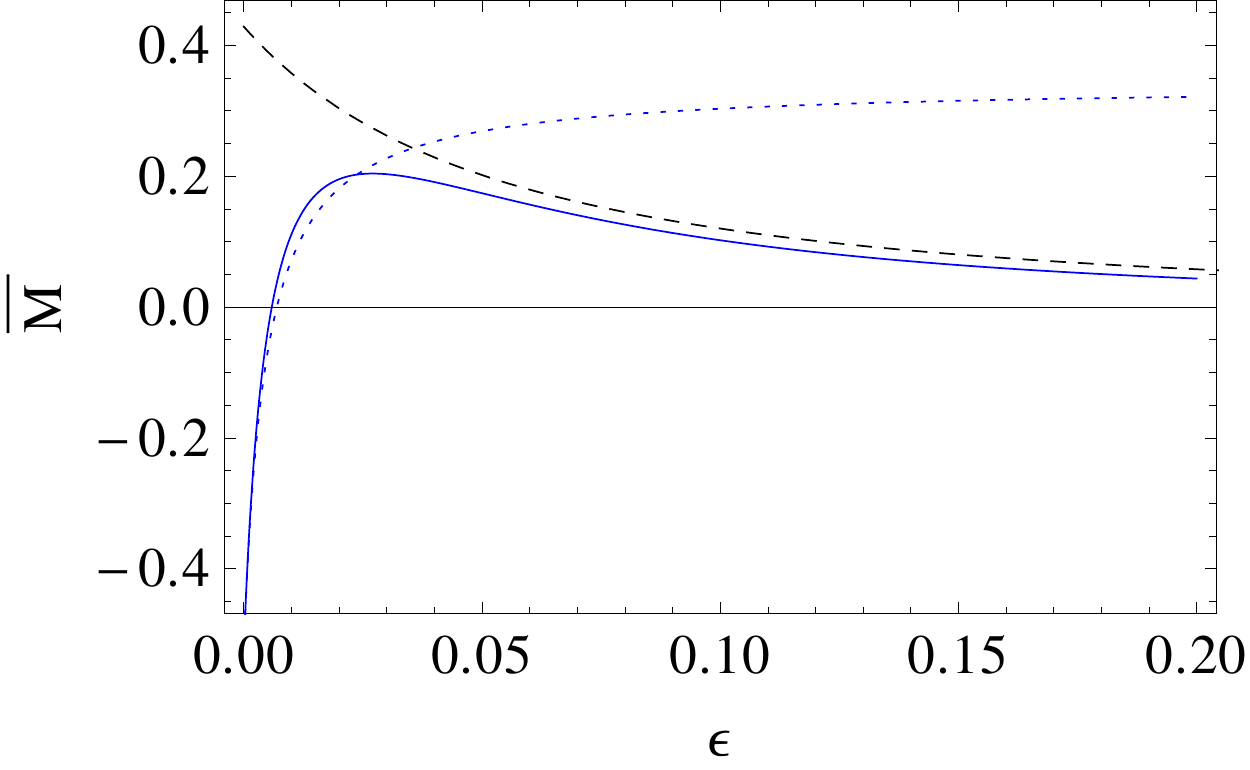}}
\put(30,15){\includegraphics[width=3cm,trim= 0cm 0cm 0cm 0cm,clip]{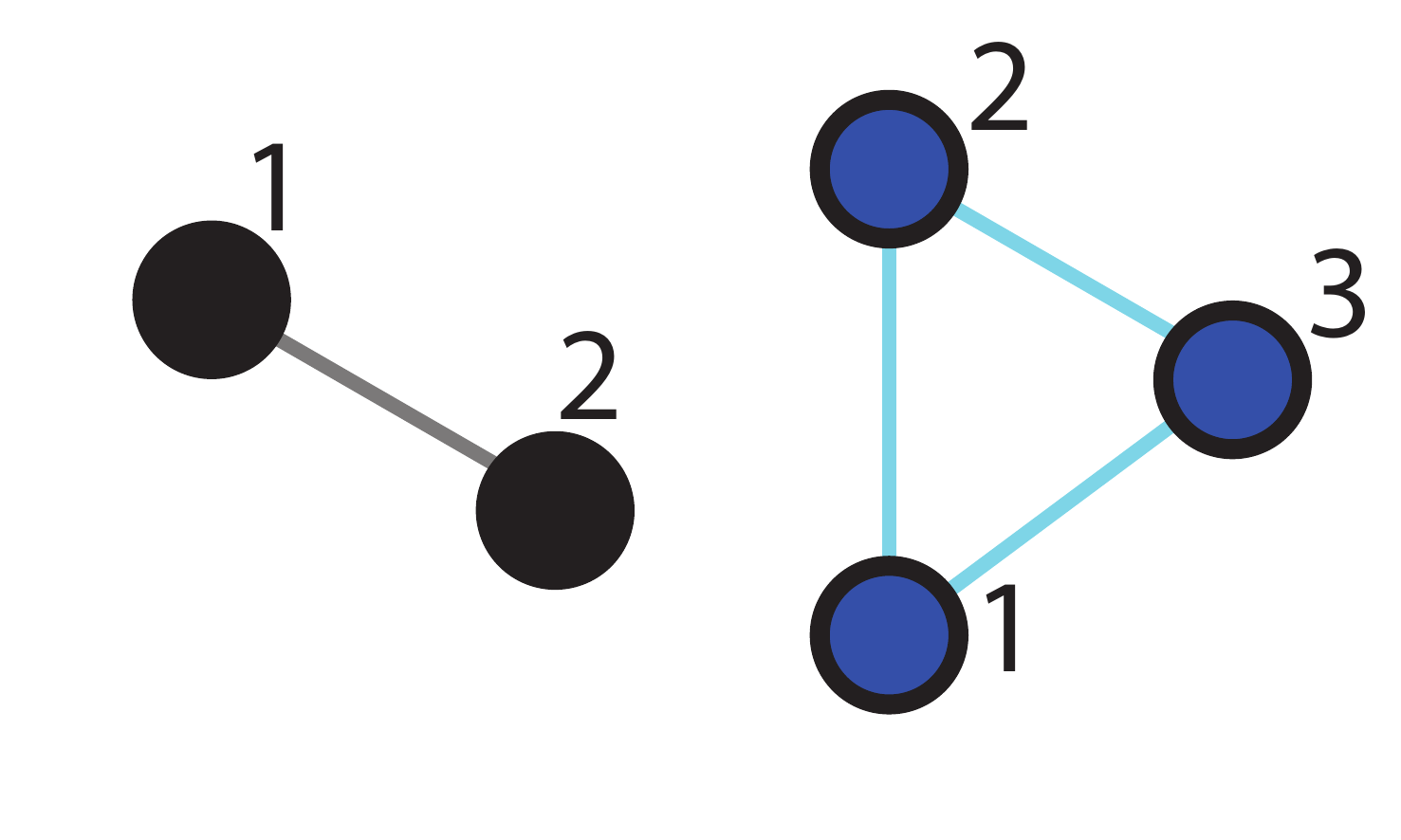}}
\put(-10,-75){\includegraphics[width=8cm,trim= 0cm 0cm 0cm 0cm,clip]{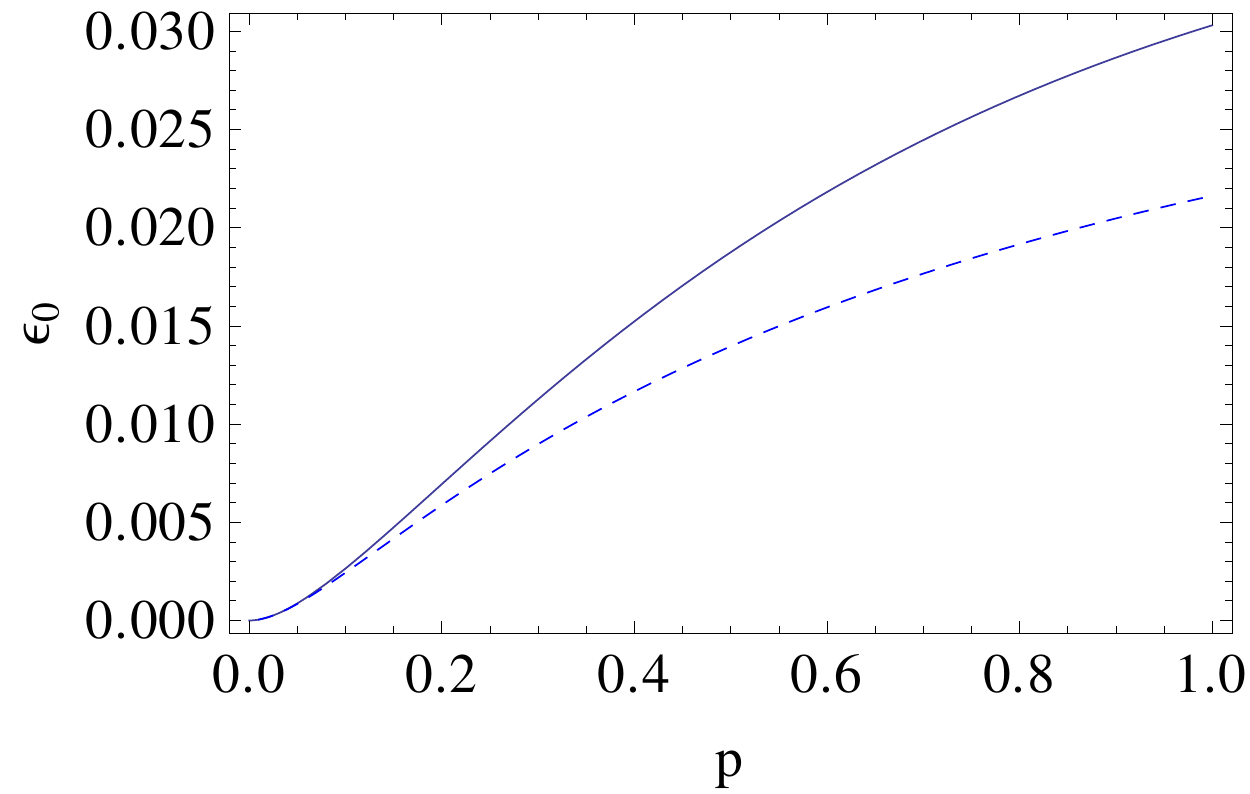}} 
\put(-10,65){{\bf a}}
\put(-10,-10){{\bf b}}
\put(20,61){$n=2$}
\put(15,20){\textcolor{blue}{$n=3$}}
\put(15,38){$\epsilon_0$}
\put(55,-40){Eq.~\ref{eq:crossing}}
\end{overpic}
\vspace{150pt}
\caption{ {\bf Finite system.}
{\bf a}, Zero crossing of $\bar{M}$ for $p=.2$. 
Dashed black curve shows the analytical solution for $n=2$, where no zero crossing is observed.
Dashed blue curve shows the analytical approximation for $n=3$ (neglecting nonlinear terms in $\varepsilon$).
Solid blue curve shows the full solution for $n=3$, where all powers of $\varepsilon$ were retained.
The zero-crossing for $n=3$ is indicated by the label $\varepsilon_0$.
{\bf b}, The level of noise $\varepsilon$ where a zero crossing for ``magnetization'' $\bar{M}$ occurs for different values of $p$.
Dashed and solid blue curves again show analytical approximation (Eq.~\ref{eq:crossing}) and numerical solution, respectively. 
}
\label{fig:3site}
\end{center}
\end{figure}

\begin{figure}[!]
\begin{center}
\begin{overpic}[width=5cm,angle=-90,trim= 0cm 0pt 0pt 0pt,clip]{dummy.pdf}
\put(-13,-18){\includegraphics[height=5.8cm,trim= 0cm 0cm 0cm 0cm,clip]{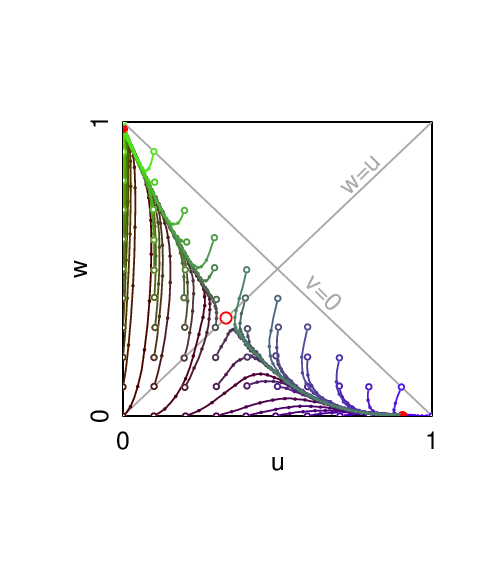}}
\put(53,-18){\includegraphics[height=5.8cm,trim= 2cm 0cm 0cm 0cm,clip]{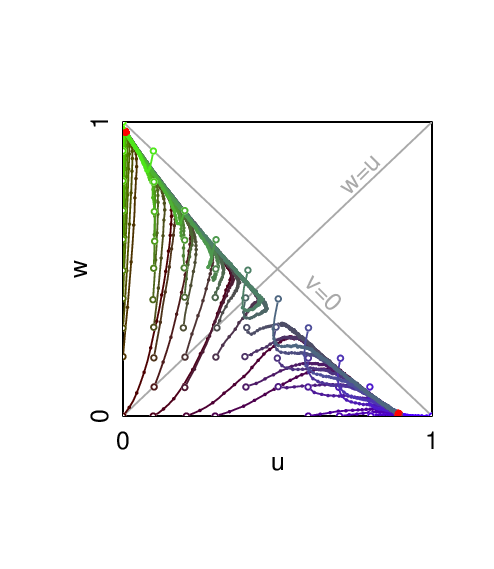}}
\put(5,47){{\bf a}}
\put(54,47){{\bf b}}
\end{overpic}
\vspace{0pt}
\caption{{\bf Phase portraits for fully and sparsely-connected systems.}
{\bf a}, Example of phase portrait in $w$-$u$-plane for the fully-connected system (Fig.~2a, main text) for $p=.2$ and $\varepsilon=.02$, i.e. for the bistable phase. 
Gray open/red filled/red open circles represent initial conditions/stable fixed points/unstable fixed points. 
Colors of curves distinguish trajectories.
{\bf b}, Similar to (a) but for a regular random graph with $\langle k\rangle=3$, Fig.~2b, main text.
Panels were obtained using simulations with $30000$ nodes. }
\label{fig:trajectory_comparison}
\end{center}
\end{figure}

\subsection{Finite systems}\label{sec:finite}
For small systems consisting only of few sites and links connecting each pair of sites, the steady state can be computed exactly.
As each node can take one of three states, the $n$-particle state space consists of $3^n$ states, i.e. the basis increases exponentially with the number of sites. 
However, considering permutation symmetry the basis reduces to $m(n)=(n^2+3n+2)/2$ states, i.e. the basis then only increases approximately quadratically with the number of sites.

The smallest interacting system is that consisting of only two nodes, connected by a link.
Hence, there are $m(2)=6$ basis states.
By evaluating all $m(2)\times m(2)$ transition probabilities $P(\chi_j\rightarrow \chi_i)$ for transitions between $n$-particle states $\chi_i$ and $\chi_j$ with $i,j\in\{1,\dots,m\}$ we write a dynamical equation for the occupation probabilities $\Pi_i$:
\begin{eqnarray}
 \dot{\Pi}_{i}&=&\sum_{j=1}^{n}\left(\Pi_{j}P(\chi_j\rightarrow \chi_i)-\Pi_{i}P(\chi_i\rightarrow \chi_j)\right)\;.
 \label{eq:3site}
\end{eqnarray}
Solving for $\Pi_i=0$ and considering the vector ${\bf{M}}\equiv (M_1,\dots,M_m)$ of ``magnetization'', where $M_i\equiv w_i-u_i$ for each $n$-particle state, we obtain the average ``magnetization'' $\bar{M}\equiv \sum_i\Pi_iM_i$ as a function of noise $\varepsilon$ and the asymmetry parameter $p$, $\bar{M}=\bar{M}(\varepsilon,p)$.
As a function of $\varepsilon$, for any given value of $p\in [0,1]$ we find that, when $n=2$, $\bar{M}$ maintains its sign for all values of noise $\varepsilon$ (Fig.~\ref{fig:3site}a).
Conversely, for $n=3$, $\bar{M}$ has a zero-crossing as $\varepsilon$ is increased.
Retaining only terms linear in $\varepsilon$ we approximate the curve $\bar{M}(\varepsilon,p)$ and,  by setting $\bar{M}(\varepsilon,p)=0$, reach an approximate expression for the zero crossing $\varepsilon_0(p)$ as a function of $p$:
\begin{equation}
\varepsilon_0(p)=\frac{p ^2 (1+p )}{2+19 p +24 p ^2+19 p ^3+2 p ^4}\;.
\label{eq:crossing}
\end{equation}
We find that the function $\varepsilon_0(p)$ increases monotonically with $p$ (Fig.~\ref{fig:3site}b) and approaches a finite value $\varepsilon_0(1)=1/33$ as $p\rightarrow 1, p<1$.
Hence, even infinitesimal asymmetry $p\neq 1$ in the model (Fig.~1) can induce a noise-dependent change of sign akin to the original Parrondo paradox.

Examining larger system ($n>3$) numerically, we find the zero crossing and qualitative behavior of $\bar{M}$ to be similar to that of $n=3$.
However, when considering a comparison with the noise effect in the Parrondo paradox, as system size is increased, the system at hand will hardly reside in a state near $\bar{M}$.
This is because the model (Fig.~1) induces bistability, whereas this effect is absent in the Parrondo paradox (discussed in the main text).

\subsection{Infinite system}\label{sec:fully_connected_graph}
To see the existence of bistability, consider the limit $n\rightarrow \infty$, where each node is connected with any other. 
In this limit, the probability for one node to be influenced by any given neighbor in two consecutive node updates decays as $1/n$. 
For sufficiently large $n$ it hence becomes appropriate to neglect all correlations and assume any node to only ``feel'', i.e. respond to, the mean densities $\bar{u}$, $\bar{v}$, and $\bar{w}$.
With $\bar{u}+\bar{v}+\bar{w}=1$ one is left with two coupled nonlinear differential equations describing $d\bar{u}/dt$ and $d\bar{w}/dt$.

\begin{figure}[ht!]
\begin{center}
\begin{overpic}[width=5cm,angle=-90,trim= 0cm 0pt 0pt 0pt,clip]{dummy.pdf}
\put(-10,20){\includegraphics[width=8cm,trim= 0cm 2.5cm 0cm 0cm,clip]{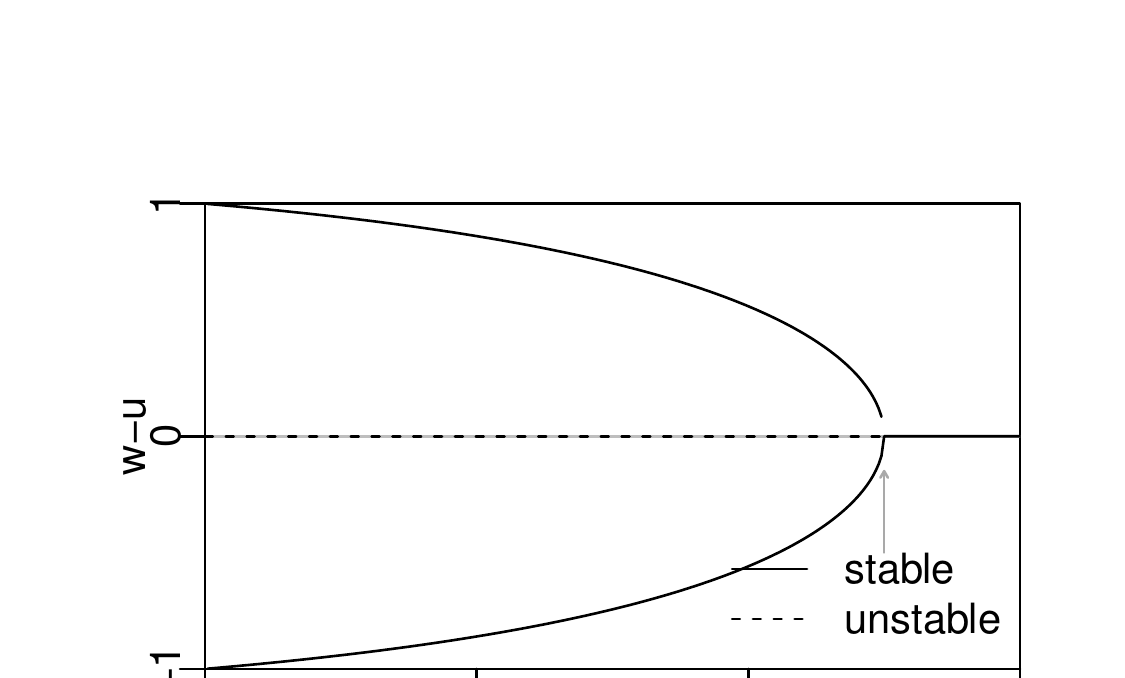}}
\put(-10,-55){\includegraphics[width=8cm,trim= 0cm 0cm 0cm 0cm,clip]{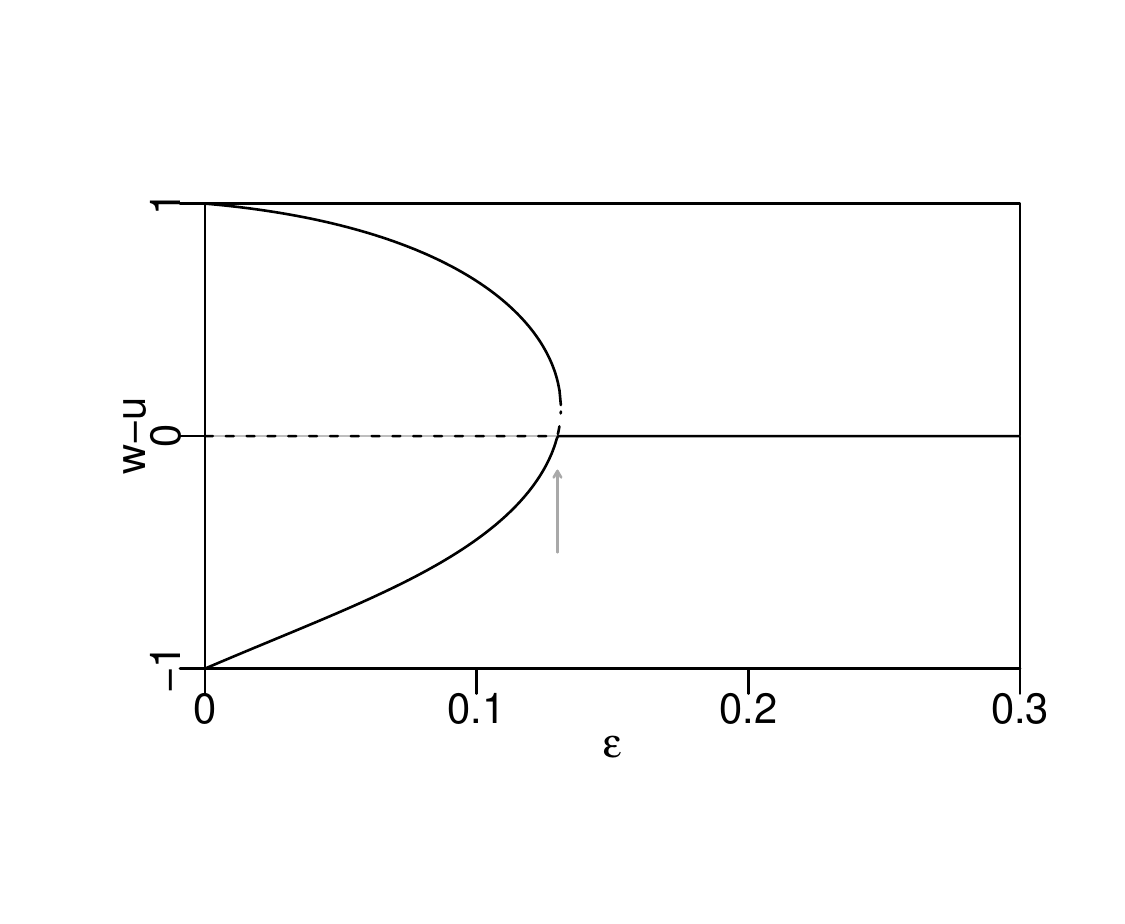}}
\put(-2,67){{\bf a}}
\put(-2,17){{\bf b}}
\end{overpic}
\vspace{70pt}
\caption{ {\bf Dynamics for infinite fully-connected system.}  
Fully connected random graph, i.e. $k=N-1$ for all nodes.
{\bf a---b}, Similar to Fig.~2 but for the fully connected graph. 
Linearly stable and unstable fixed points are shown as solid and dashed lines, respectively.
Note the absence of a transition from one polarized state to another in (a) and (b). 
}
\label{fig:fully_connected_graph}
\end{center}
\end{figure}

We now show that these equations allow for two stable fixed points where $\bar{M}\neq 0$ when $\varepsilon<\varepsilon_c(p)$. 
For $\varepsilon\geq\varepsilon_c(p)$ a stable fixed point exists at $\bar{M}=0$.
This is the extreme case of the model in Fig.~1 (main text) where the limit of large systems and full connectivity is considered, i.e. $N\rightarrow \infty$ and $2L/N=N-1$.
As mentioned, when all nodes are connected any correlation between individual nodes is lost. 
The conditional probabilities in Fig.~1b can then just be described as contingent on mean densities of the three states $\bar{u}$, $\bar{v}$, and $\bar{w}$.
We drop the overbar for simplified notation. The resulting dynamical equations then read
\begin{eqnarray}
 \dot{u}&=&f(u,w)\;,\label{eq:f}\label{eq:dotU}\\
 \dot{w}&=&g(u,w)\label{eq:g}\label{eq:dotW}\;.
\end{eqnarray}
The functions $f(u,w)$ and $g(u,w)$ depend only on the densities of the states $u$ and $w$ because the density of $v$ results from the conservation of probability, i.e. $v(w,u)=1-w-u$.
Using the conditions in Fig.~1, the expression on the RHS of Eqs~\ref{eq:dotU} and \ref{eq:dotW} are:
\begin{widetext}
\begin{eqnarray}
 f(u,w)&=&(1-\varepsilon)u\left[p_2h(w,u)-p_3w \right]+\varepsilon(h(w,u)-u)=(1-\varepsilon)u\left[p_2-(p_2+p_3)w-p_2u \right]+\varepsilon(1-w-2u)\label{eq:f}\\
 g(u,w)&=&(1-\varepsilon)w\left[p_4h(w,u)-p_1u \right]+\varepsilon(h(w,u)-w)=(1-\varepsilon)w\left[p_4-(p_4+p_1)u-p_4w \right]+\varepsilon(1-2w-u)\;.
 \label{eq:g}
\end{eqnarray}
\end{widetext}
To further discuss the fixed points of Eqs~\ref{eq:f} and \ref{eq:g}, as in the main text we again specialize to the case where $p_2=p_3$ and $p_4=p_1$. 
By rescaling time, we can always express all quantities in units of $p_1$, i.e. we make the replacement $p\equiv p_2\rightarrow p_2/p_1$, $\varepsilon\rightarrow\varepsilon/p_1$ and $p_1\rightarrow 1$.
The equations \ref{eq:f} and \ref{eq:g} then read:
\begin{eqnarray}
 f(u,w)&=&(1-\varepsilon)pu\left(1-2w-u \right)+\varepsilon(1-w-2u)\\
 g(u,w)&=&(1-\varepsilon)w\left(1-2u-w \right)+\varepsilon(1-2w-u)\;.
\end{eqnarray}
For these simplified equations, there is a line of fixed points at $u=w=1/3$ for all $\varepsilon$ and linear stability analysis of these fixed points reveals a bifurcation point at
\begin{equation}
 \varepsilon_c=\left(1+3/\sqrt{p}\right)^{-1}\;.
 \label{eq:eps_c}
\end{equation}
For $\varepsilon<\varepsilon_c$ this fixed point is an unstable saddle, while for $\varepsilon>\varepsilon_c$ it is a stable node.
The characterization of the remaining fixed points depends on the remaining parameter $p$ (Fig.~\ref{fig:fully_connected_graph}).
For $p=1$, in the diagram of $M=w-u$ two symmetric branches of stable fixed points exist for $\varepsilon<\varepsilon_c=1/4$, which approach $\pm 1$ as $\varepsilon\rightarrow 0$.
In this case, solving for $f(u,w)=g(u,w)=0$ yields the fixed points 
\begin{eqnarray}
 u&=&\frac{1}{2}\left(1-\varepsilon'\pm \sqrt{1-2\varepsilon'-3\varepsilon'^2}\right) \label{eq:branches1}\\
 w&=&\frac{1}{2}\left(1-\varepsilon'\mp \sqrt{1-2\varepsilon'-3\varepsilon'^2}\right)\;,
 \label{eq:branches2}
\end{eqnarray}
where we have defined $\varepsilon'\equiv\varepsilon/(1-\varepsilon)$ for simplified notation (Fig.~\ref{fig:fully_connected_graph}a and Fig.~2a, gray curve).
The difference $w-u$ becomes
\begin{equation}
M=w-u=\pm\sqrt{1-2\varepsilon'-3\varepsilon'^2}\;.
\end{equation}

For $p<1$, $\varepsilon_c$ diminishes and the branches of stable fixed points become asymmetric (Figs~2 and ~\ref{fig:fully_connected_graph}b).
An additional line of unstable fixed points appears, which joins the bifurcation point and the upper branch of stable fixed points.
Analytical solutions are hard to come by for this asymmetric case.

Consider now a system of low noise, i.e. $\varepsilon\rightarrow 0$. 
In this case, the system will approach one of the stable fixed points $w-u=\pm 1$.
As the noise level is increased, the state of the system will follow the corresponding branch of stable fixed points, until it eventually reaches $w-u=0$. 
In the case of the lower branch, i.e. $w-u<0$, this approach will be continuous. 
In the case of the upper branch, i.e. $w-u>0$, there will be a jump discontinuity (compare Fig.~\ref{fig:fully_connected_graph}b). 
However, in all cases, the neutral value of $w-u=0$ will be approached monotonically as the noise level is increased.
The effect of noise hence is to reduce the polarization of the system state and eventually yield a state where all three configurations $u$, $h$ and $w$ are equally likely.
 
Maintaining the limit of sufficiently large system size, we have confirmed that the results described in the main text for the infinite system size limit are retained when using numerical simulation for large systems ($n>10^4$ sites), hence benchmarking our quasi-infinite system size simulations with the exact results.
Lowering the average degree $\langle k\rangle$ we retrieve the effect of correlations, hence $\langle k\rangle\gtrsim 1$.
In these simulations with correlations, we refer to $\varepsilon^*(p)$ as the transition point between a single and multi-valued function $\bar{M}(\varepsilon,p)$.

\section{Sparsely connected systems --- Mean field approximation}\label{sec:extended_mf}
Consider again the network of $N$ nodes but take the connectivity to be sparse, i.e. $L/N\ll N$.
For simplicity, assume that the links are equally distributed among all nodes, so that each node has exactly $k$ connections to other nodes. 
This topological structure is sometimes called a {\it regular graph} of degree $k$.
In the following, we refer to the link between nodes $n$ and $r$ as $l_{nr}$.
Choose any node at random for update. 
Consider a central node and its $k$ nearest neighbors, forming a cluster of $k+1$ nodes and $k$ links.
In general, the configuration space of this $k+1$-node cluster spans $s_k\equiv 3^{k+1}$ states. 
We refer to any of the $k-node$ states $i$ as $\chi_i$ and to the probability of this state being occupied as $\Pi_i$. 
Writing the dynamical equations for the transition probabilities (Fig.~1, main text) for each transition between these states, the steady-state solutions can be computed. 

\begin{figure}[h!]
\begin{center}
\includegraphics[width=8cm,trim= 0cm 0cm 0cm 0cm,clip]{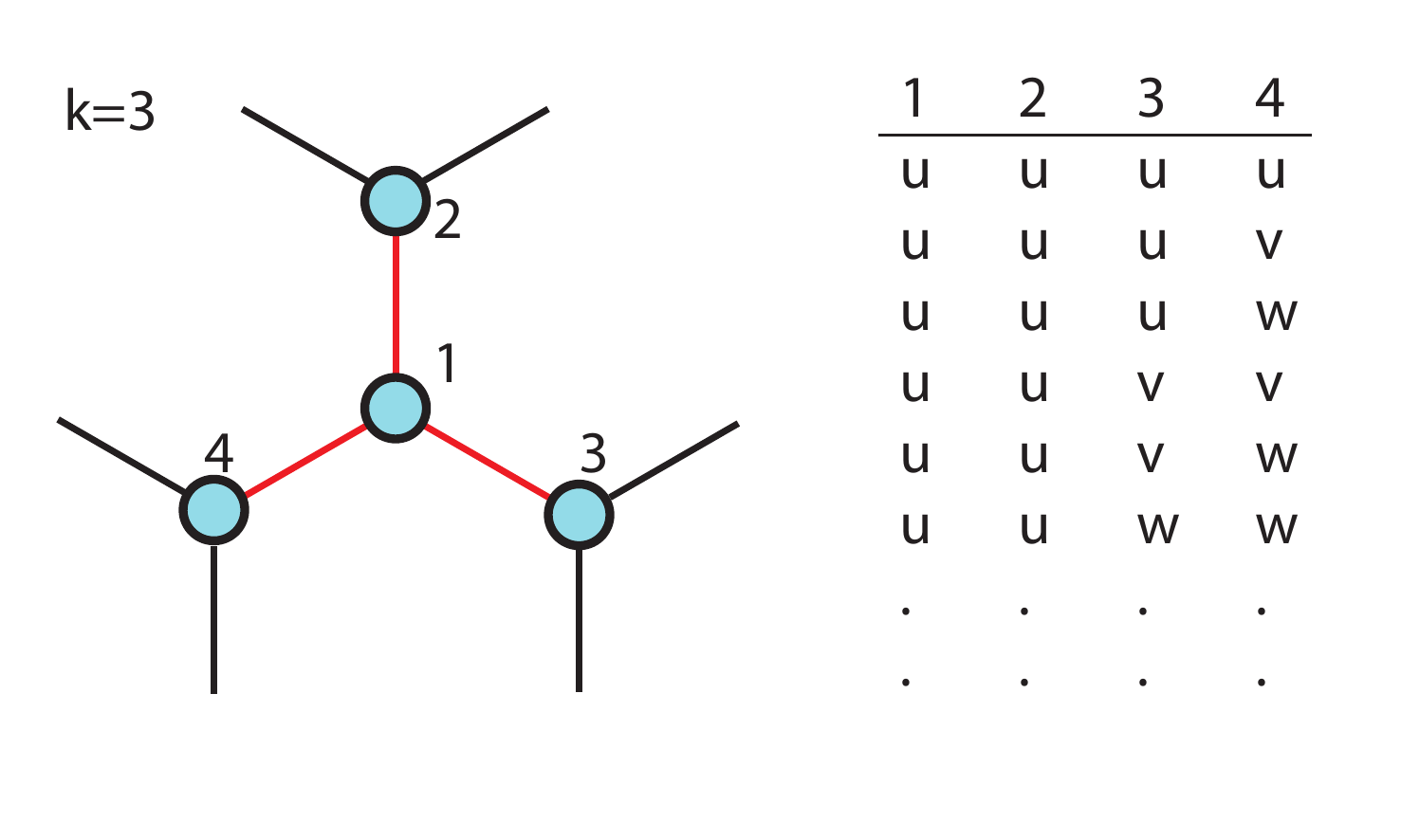}
\caption{ {\bf Mean-field approximation.}
Example for connectivity $k=3$. A central node is connected to three nearest neighbors, which themselves are each connected to three nearest neighbors. 
In the tree-approximation it is assumed that nearest neighbors are connected only by one path.
We distinguish internal and external links, which are indicated by red and black colors, respectively.
{\it Also shown}: several configurations of unique four-node states, i.e. those that do not map onto one another by permutations of the individual single-node states.
}
\label{fig:MF_schematic}
\end{center}
\end{figure}

However, this would ignore all interaction with the surroundings of the cluster. 
In order to achieve a better approximation, consider the following: When all nodes have equal connectivity, choosing a target node and its partner is equivalent to choosing a link and assigning one of the two nodes involved as the target node. 
Now consider all links {\it within} the cluster (shown in red in Fig.~\ref{fig:MF_schematic}) as well as those {\it outside} the cluster (black). 
There are $k$ ways to choose an internal link and the target node will always lie within the cluster. 
There are $k(k-1)$ ways to choose external links, however, the target node will only lie within the cluster at probability $1/2$. 
Together, the probability to choose internal updating is then 
\begin{equation}
 p_{int}=1-p_{ext}=\frac{k}{k+k(k-1)/2}=\frac{2}{1+k}\;.
\end{equation}

We now need to separate the two cases of selecting an internal versus external link. 

\noindent
{\bf Internal update ---} Selecting a link and target node generates a transition matrix for all cluster states.
A given state $\chi_i$ thereby has a probability $P(\chi_i\rightarrow \chi_j)$ to transition to state $\chi_j$.
This probability is straightforward to determine, given the schematic in Fig.~1 and noting that any possible transition is constrained to states that differ only by one elementary move. 
E.g. the transition $\{uuuv\}\rightarrow \{uuuw\}$ is possible when node four is selected as the target site and its probability is then $\varepsilon/2$.
Conversely, the transition probability $P(\{uuuv\}\rightarrow \{uuuu\})=(1-\varepsilon)p+\varepsilon/2$ since a coordinated move is now possible.

\noindent
{\bf External update ---} Updating an external link can modify any but the central node of the cluster. 
We now need to consider the updating of {\it external} cluster states, i.e. clusters with central nodes located at sites $2,\dots,k$+$1$ w.r.t. the original cluster. 
These updates require some book keeping: For a given cluster state $\chi_i$ not all external cluster states will be possible. 
A sum needs to be carried out over nodes $j\in \{2,\dots,k+1\}$, where the respective states of the pairs $(1,j)$ constrain the possible external cluster states.
Given this subset of states, all possible transitions are again enumerated. 
It is then back-tracked how these transitions affect the respective node $j$, which in turn leads to an update of the internal cluster state.

The temporal evolution, in time units of $N$ node updates, of the occupation probability  $\Pi_{i}$ hence is:
\begin{widetext}
\begin{eqnarray}
 \dot{\Pi}_{i}&=&p_{int}\left(\sum_{j=1}^{n}\Pi_{j}P(\chi_j\rightarrow \chi_i)-\Pi_{i}P(\chi_i\rightarrow \chi_j)\right)\nonumber\\ 
 &+&(1-p_{int})  \sum_{x=2}^{k+1}\left(\sum_{j}
 \Pi_{j}P(\{s_q\}|s_1,s_x')P(\chi_{s_1,s_x',\{q\}}\rightarrow \chi_{s_1,s_x,\{q\}})
 -\Pi_{i}P(\{s_q\}|s_1,s_x)P(\chi_{s_1,s_x,\{q\}}\rightarrow \chi_{s_1,s_x',\{q\}})
 \right)\;.
 \label{eq:mean_field_links}
\end{eqnarray}
\end{widetext}
Here, 
\begin{equation}
P(\{s_q\}|s_1,s_x)\equiv \frac{\Pi_{1,x,\{q\}}}{\sum_{\{q\}}\Pi_{1,x,\{q\}}}
\label{eq:conditional_prob}
\end{equation}
is the conditional probability that the remaining nodes $\{q\}$ are in the states $\{s_q\}$ given that nodes $1$ and $x$ are in the single-node state $s_1$ and $s_x$, respectively. 
The sum in the denominator extends over all configurations of external states that are compatible with the single node states at sites 1 and $x$ and serves as a normalization.

Notably, in Eq.~\ref{eq:mean_field_links} in the first term all configurations $\chi_i$ of the internal cluster can contribute while in the second term only those configurations $\Pi_j$ contribute, which are compatible with a given single-node state $s_1$ at the central site and a state $s_x$ at a given peripheral site.
Note further that in the first term only linear contributions in the occupation probabilities $\Pi_i$ occur, while the second term involves also product terms of probabilities $\Pi_i$.

The mean field approximation in the equation consists in the closure of correlations up to the size of the cluster. 
It is implicitly assumed that longer-ranged correlations do not contribute and can be neglected.
Eq.~\ref{eq:mean_field_links} constitutes a set of non-linear equations in the occupation probabilities $\Pi_i$, which itself has to be solved numerically.

\begin{figure*}[ht!]
\begin{center}
\begin{overpic}[width=5cm,angle=-90,trim= 0cm 0pt 0pt 0pt,clip]{dummy.pdf}
\put(-45,-92){\includegraphics[width=6.6cm,trim= .7cm 0cm 0cm 0cm,clip]{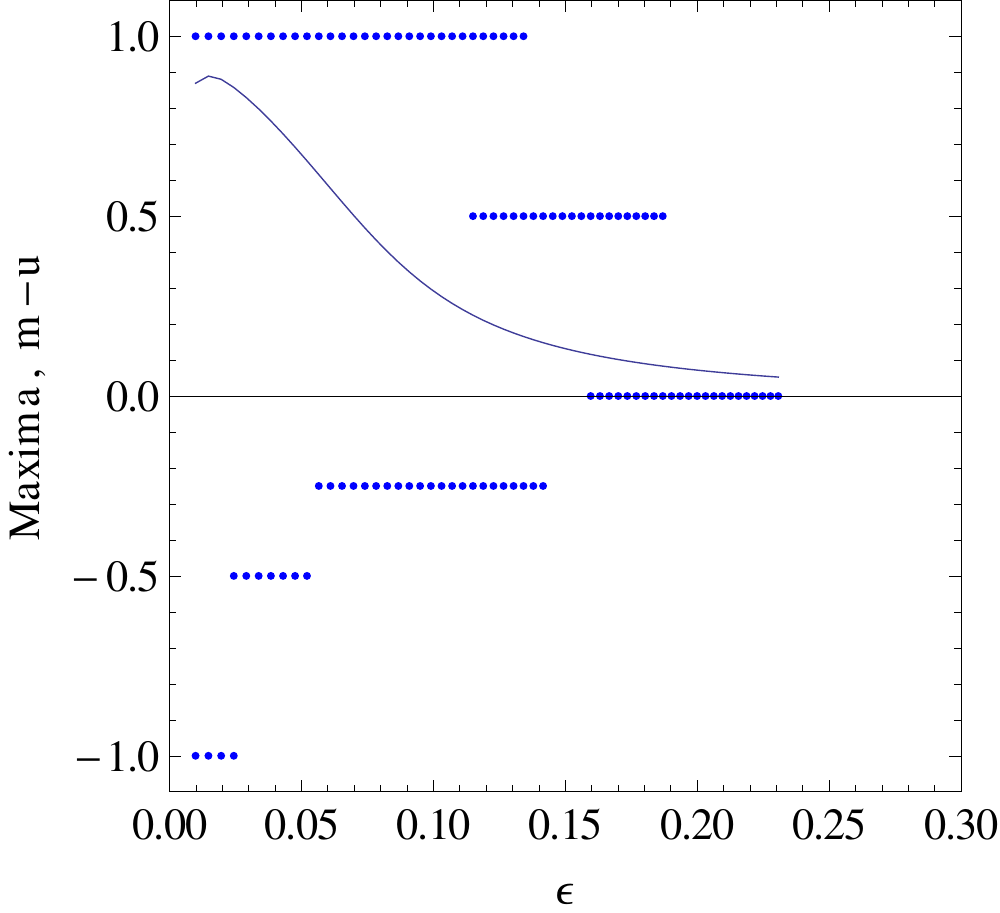}}
\put(-45,0){\includegraphics[width=6.6cm,trim= .7cm 1.2cm 0cm 0cm,clip]{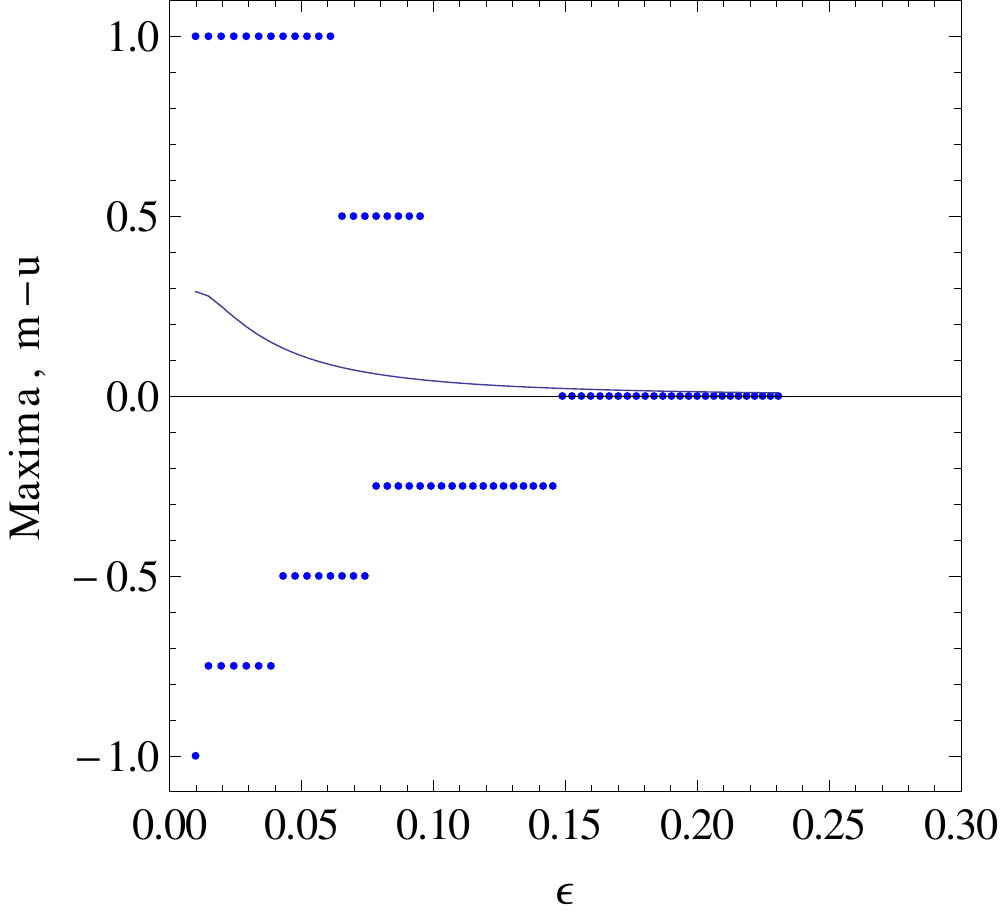}}
\put(55,-92){\includegraphics[width=5.84cm,trim= 1.7cm 0cm 0cm 0cm,clip]{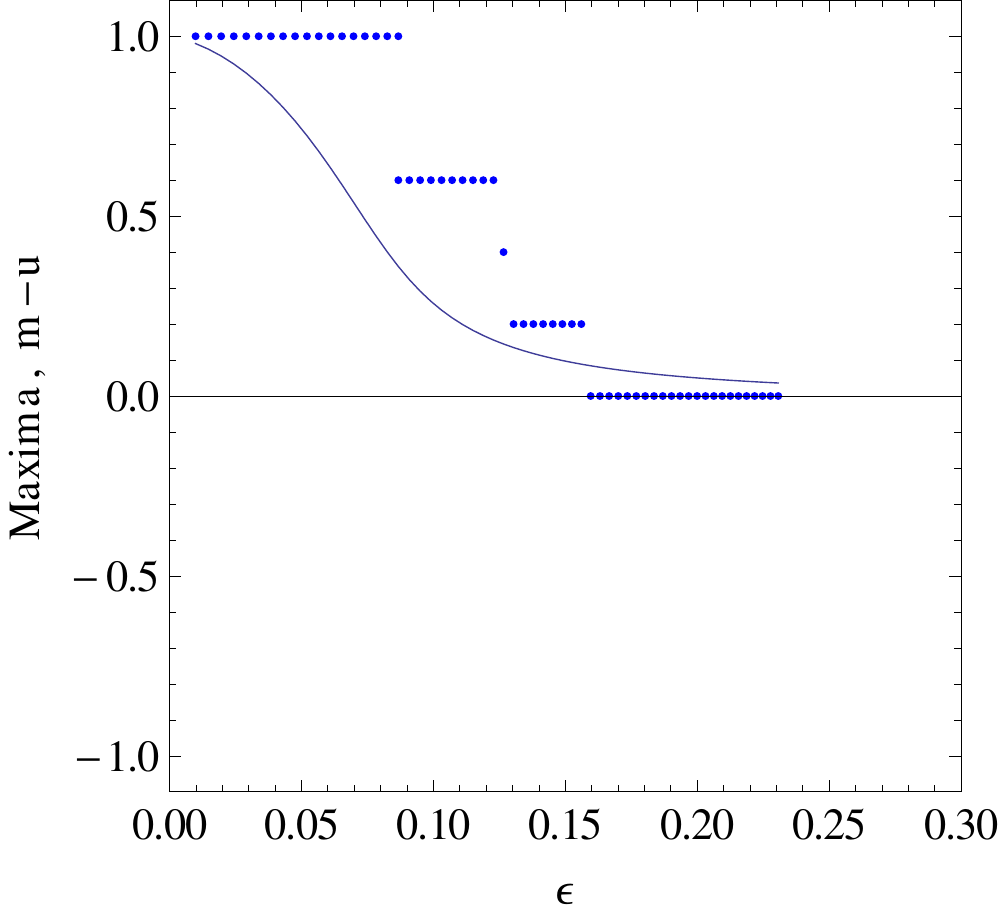}}
\put(55,0){\includegraphics[width=5.84cm,trim= 1.7cm 1.2cm 0cm 0cm,clip]{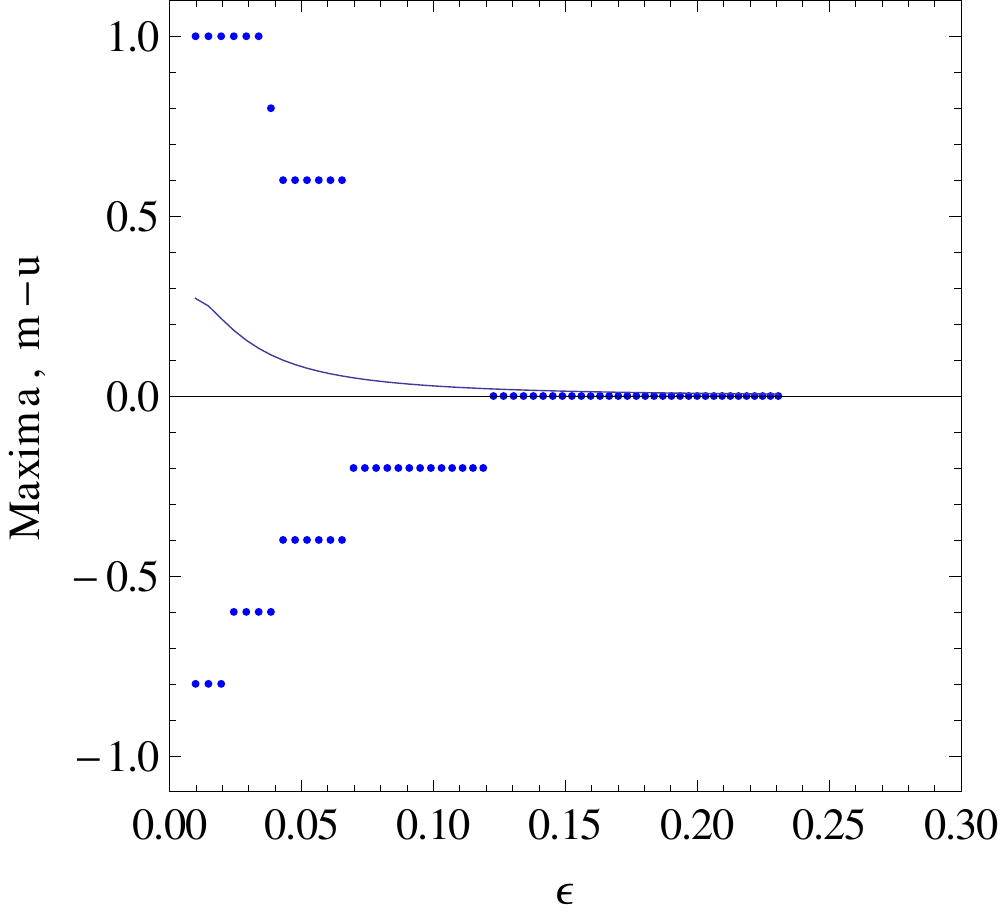}}

\put(37,73){{\bf a}}
\put(37,-7){{\bf b}}
\put(125,73){{\bf c}}
\put(125,-7){{\bf d}}

\put(-50,-20){\rotatebox{90}{Maxima of w-u}}

\end{overpic}
\vspace{175pt}
\caption{ {\bf Maxima of probability distribution function.} 
{\bf a}, Star topology consisting of four sites, as shown schematically in Fig.~\ref{fig:MF_schematic} as red links.
Solid points show the respective local maxima obtained by inspecting the density of states in the mean field solution.
Thin line shows the average value of $w-u$.
{\bf b}, As in (a) but interacting with surrounding sites, i.e. as in Fig.~\ref{fig:MF_schematic} including the black links. 
Note the more abrupt decrease of the $u$-dominant state, i.e. maxima for $w$-dominant states persist in noise regions where $u$-dominant states are absent. 
Note the increase of the average $w-u$ w.r.t. the star topology.
{\bf c}, Similar to (a) but for $N=5$ sites. 
{\bf d}, Similar to (c) but for $N=5$ sites. Note that the maxima at $u>w$ have now disappeared.
}
\label{fig:5_star_comparison}
\end{center}
\end{figure*}

Fig.~\ref{fig:5_star} shows examples of the density of states for the solution to Eq.~\ref{eq:mean_field_links} when only the internal cluster is considered. 
The density of states is initially peaked at low and high values of $w-u$, with the high-$u$ state the most likely.
As noise is increased, one peak gradually disappears and a single-peak distribution results.

The crucial features of this plot can be represented more compactly by extracting the peaks of the distribution function (Fig.~\ref{fig:5_star_comparison}). 
For the cluster consisting of four sites, excluding the surroundings, Fig.~\ref{fig:5_star_comparison}a shows the dependence of maxima on the noise $\varepsilon$.
Indeed, at low $\varepsilon$ a bimodal distribution is present, whereas for larger noise ($\varepsilon>.15$) the distribution is unimodal.
We compare this to the mean field solution where also the surroundings are taken into account (Fig.~\ref{fig:5_star_comparison}b). 
Now the peaks for the $u$-dominant state decay more rapidly, the pattern is much more skewed than in Fig.~\ref{fig:5_star_comparison}a.
At a value of $\varepsilon\approx .04$ the distribution is unimodal and peaks are present for the $w$-dominant state.
This result should be compared to the simulation result in Fig.~2b (main text), where similar qualitative features are present.

\section{Additional analysis and simulations}\label{sec:addition_simulations}

\subsection{Additional network structures}\label{sec:additional_network}
\begin{figure}[!]
\begin{center}
\begin{overpic}[width=5cm,angle=-90,trim= 0cm 0pt 0pt 0pt,clip]{dummy.pdf}
\put(-15,15){\includegraphics[width=9.5cm,trim= 0cm 2.4cm 0cm 0cm,clip]{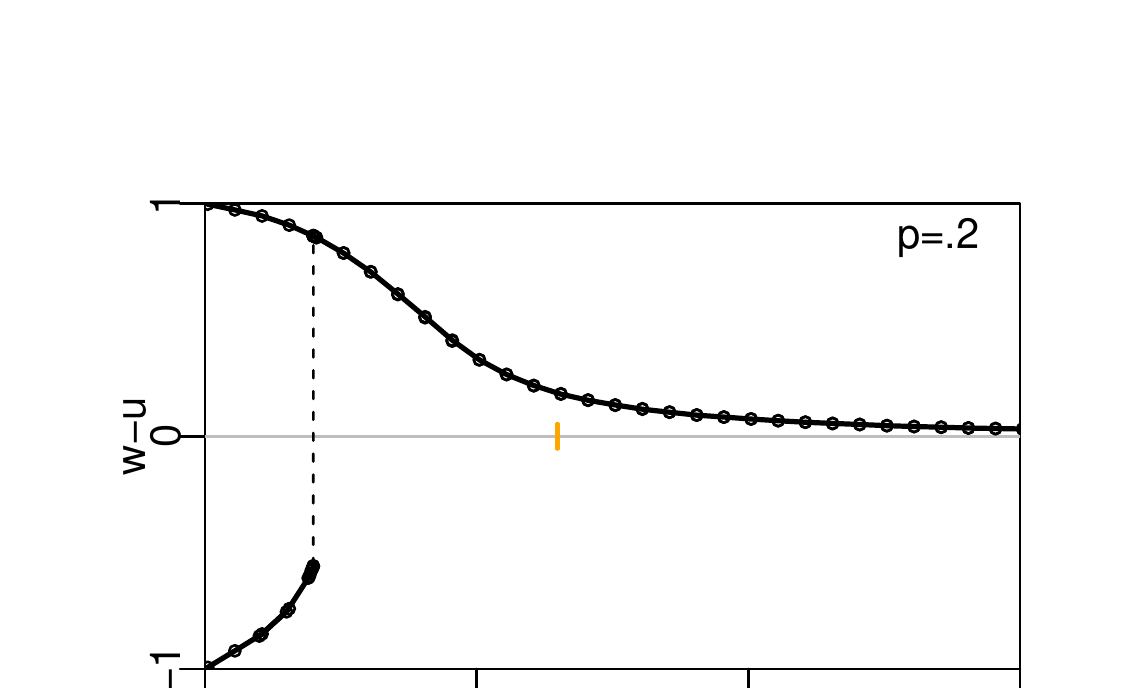}}
\put(-15,-75){\includegraphics[width=9.5cm,trim= 0cm 0cm 0cm 0cm,clip]{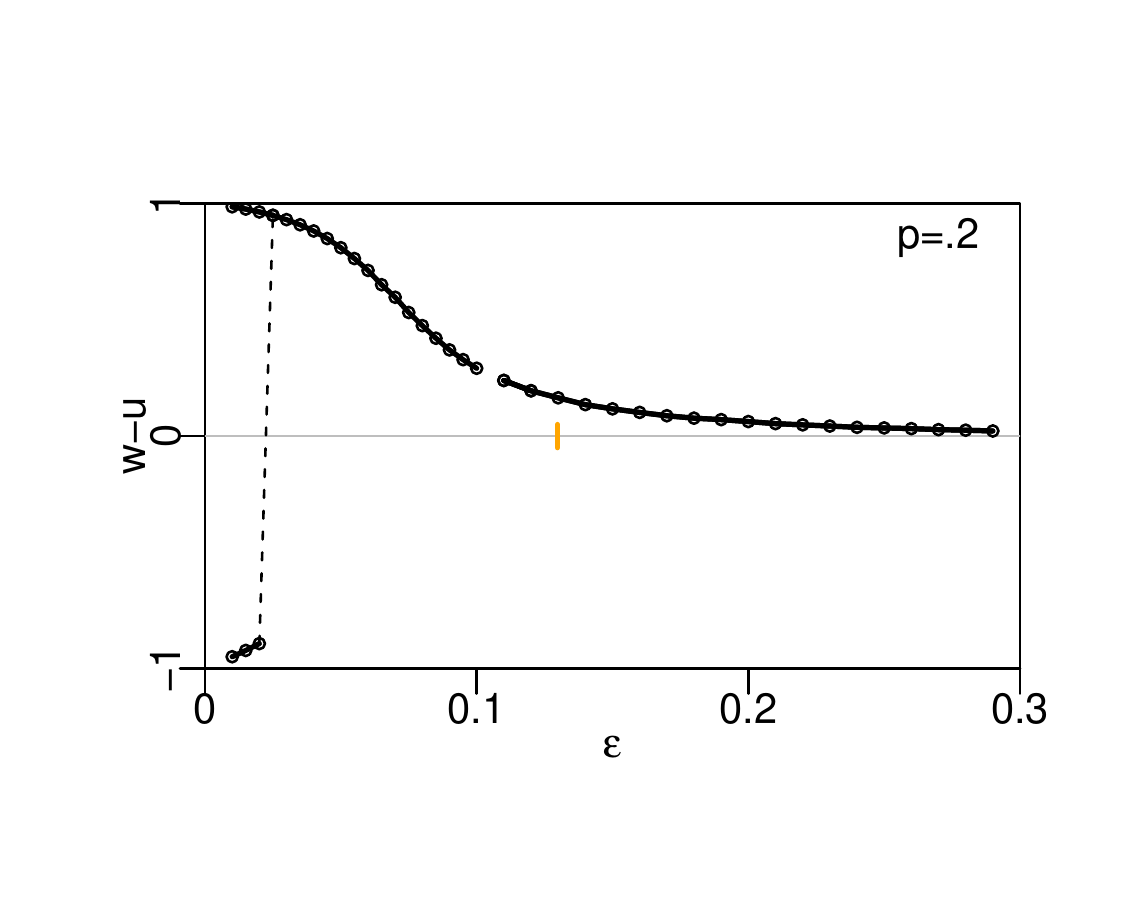}}
\put(12,60){\bf a}
\put(12, 0){\bf b}
\end{overpic}
\vspace{110pt}
\caption{{\bf Transitions for additional network structures.} 
Similar to Fig.~2c ($N=5,000$) but for a Poissonian degree distribution with $\langle k\rangle=3$ (shown in {\bf a}) as well as a square lattice (shown in {\bf b}).
}
\label{fig:transition_poisson}
\end{center}
\end{figure}

In Fig.~\ref{fig:transition_poisson} we supplement main text Fig.~2 by including two additional network geometries. 
A random graph is one where links are assigned randomly without restrictions to each node's degree. 
The resulting degree distribution is then binomial and for sufficiently large systems approximated as a Poisson distribution. 
We simulated the dynamics (Fig.~1, main text) for otherwise similar parameter configurations as in Fig.~2b,c (main text), i.e. $\langle k\rangle=3$ and $p=.2$.
The result is very similar to that of the degree-regular graph (Fig.~2b).
This means that the relatively small degree fluctuations about the average do not lead to strong deviations of the polarization.

We also simulated a two-dimensional square lattice, i.e. translationally invariant graph of degree $\langle k \rangle=k=4$ (Fig.~\ref{fig:transition_poisson}b).
Despite the larger degree, which would be expected to shift the zero-crossing to larger values of $\varepsilon$, the zero-crossing now actually moves to markedly smaller values of $\varepsilon$ as compared to the Poissonian (Fig.~\ref{fig:transition_poisson}a), degree-regular (Fig.~2b) or scale-free (Fig.~2c) cases.
This finding again highlights the potential importance of disorder in granting a large range of bistability.

\begin{figure}[ht!]
\begin{center}
\begin{overpic}[width=5cm,angle=-90,trim= 0cm 0pt 0pt 0pt,clip]{dummy.pdf}
\put(17,35){\includegraphics[width=5cm,trim= 0cm 0cm 0cm 0cm,clip]{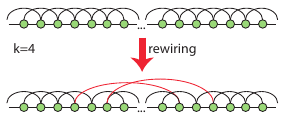}}
\put(0,-9){\includegraphics[width=7cm,trim= 0cm 2.5cm 0cm 0cm,clip]{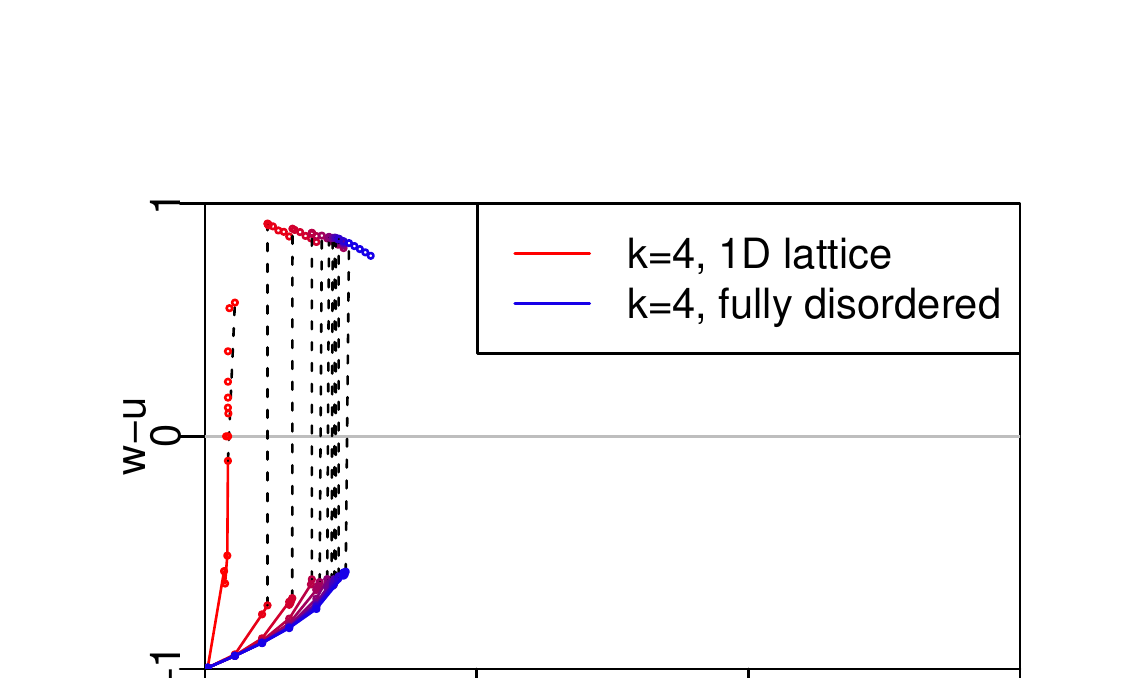}}
\put(0,-53){\includegraphics[width=7cm,trim= 0cm 2.5cm 0cm 0cm,clip]{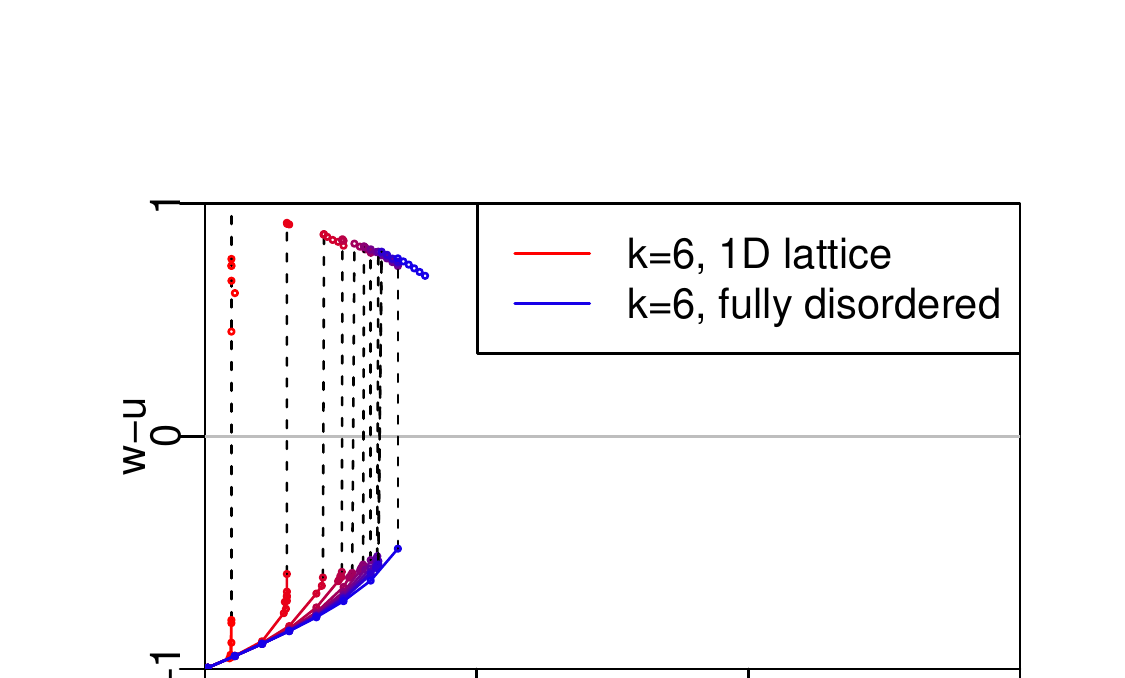}}
\put(0,-118){\includegraphics[width=7cm,trim= 0cm 0cm 0cm 0cm,clip]{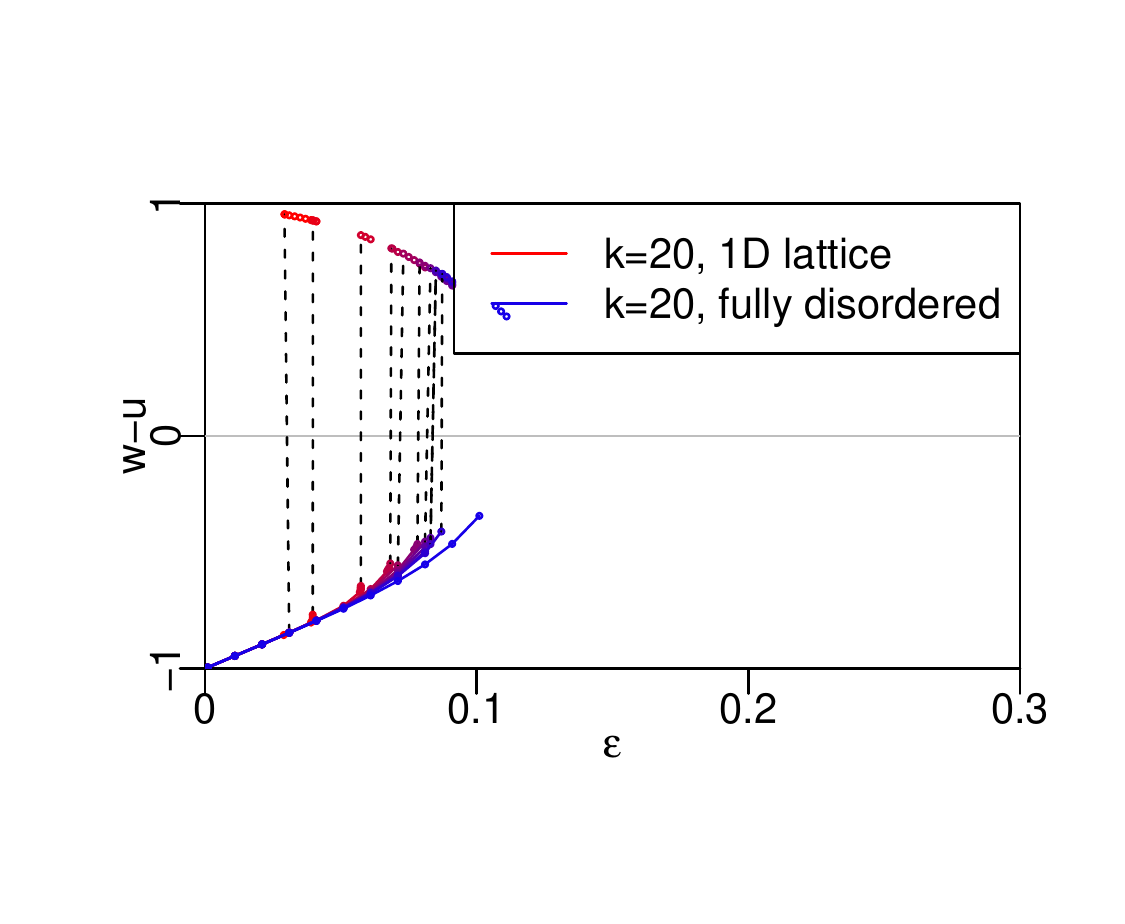}}
\put(0,-188){\includegraphics[width=7cm,trim= 0cm 0cm 0cm 0cm,clip]{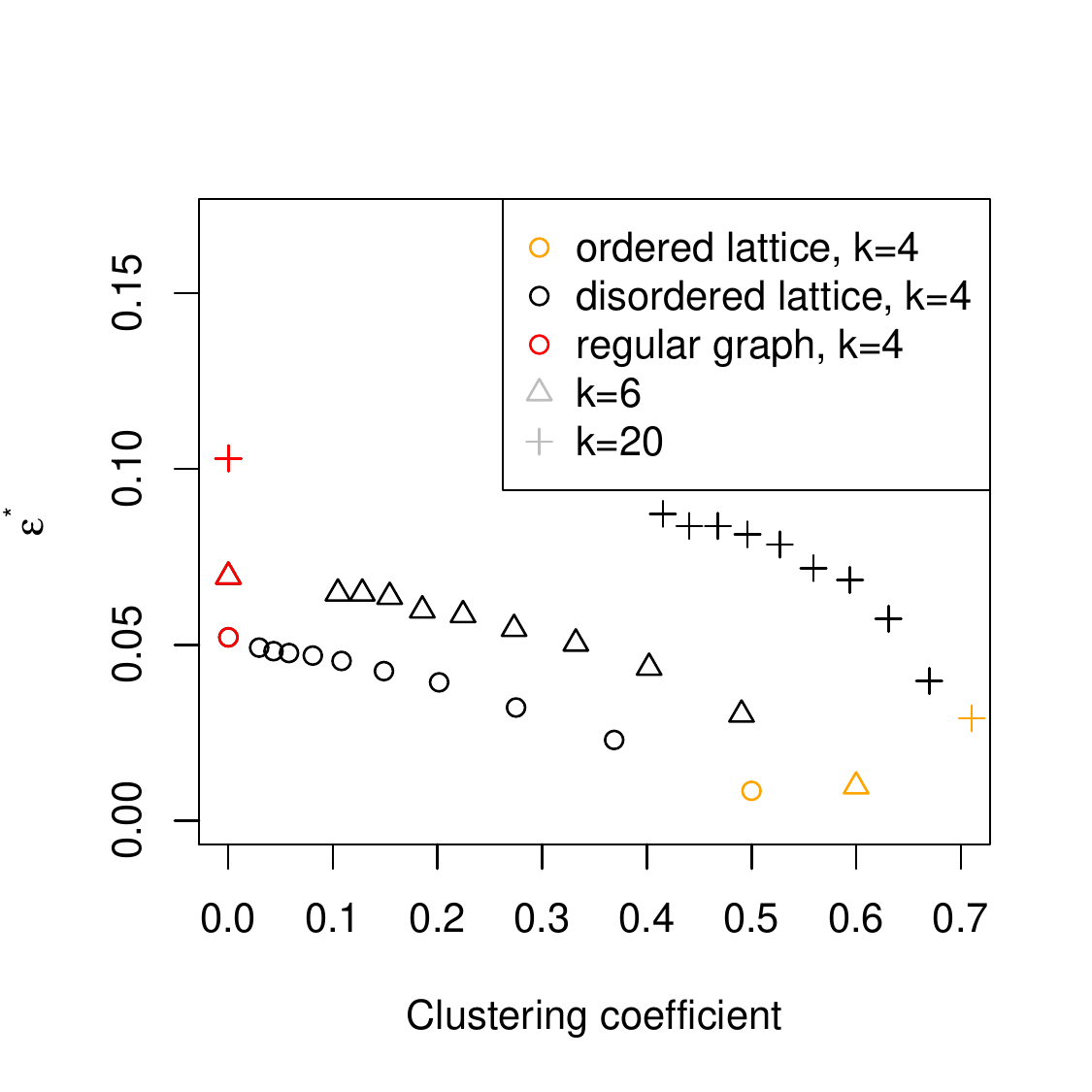}}

\put(5,60){{\bf a}}
\put(5,30){{\bf b}}
\put(5,-15){{\bf c}}
\put(5,-60){{\bf d}}
\put(5,-110){{\bf e}}
\end{overpic}
\vspace{370pt}
\caption{ {\bf Effect of clustering.}  
{\bf a}, Rewiring operation, by which a 1D-lattice with nearest and next-nearest neighbor couplings is gradually converted to a random graph of unchanged degree. 
Green points are network nodes, black lines are links, red lines indicate a pair of rewired links.
The lattice is taken to have cyclic boundary conditions.
{\bf b}, Transitions for varying levels of disorder for $k=4$, $\varepsilon^*$ increases with disorder.
The fully disordered system is a regular graph of degree $k=4$.
System size: $N=2,000$.
{\bf c}, Similar to (b) but for $k=6$. 
System size: $N=2,000$.
{\bf d}, Similar to (b) but for $k=20$.
System size: $N=5,000$.
Note that we use a larger system size for the higher connectivity here to ensure that finite size effects are still minimal.
{\bf e}, Summary of $\varepsilon^*$ for the simulations in (b)---(d) as a function of clustering coefficient.
}
\label{fig:rewiring}
\end{center}
\end{figure}

To explore this further, we tested for the effect of disordering an initially clustered system (Fig.~\ref{fig:rewiring}).
We start from a one-dimensional ring of $N$ sites and connected each node to several of its immediate neighbors by a link.
After computing the transition for this configuration, we rewired the network by swapping links randomly. 
In the link swapping procedure, two links are chosen at random, and if they do not involve any nodes twice, one end of each is exchanged with the corresponding node for the partner link \cite{maslov2002specificity}.
The effect of repeated rewiring can be captured in terms of the clustering coefficient (Fig.~\ref{fig:rewiring}). 
We find that for several different values of mean degree $\langle k\rangle$ ranging from 4 to 20, lower clustering, i.e. more disorder, means larger values of $\varepsilon^*$.

\subsection{Sensitivity analysis}\label{sec:sensitivity}
\noindent
We performed sensitivity tests regarding system size and dependence on degree. 

\begin{figure}[ht!]
\begin{center}
\includegraphics[width=8cm,trim= 0cm 0cm 0cm 0cm,clip]{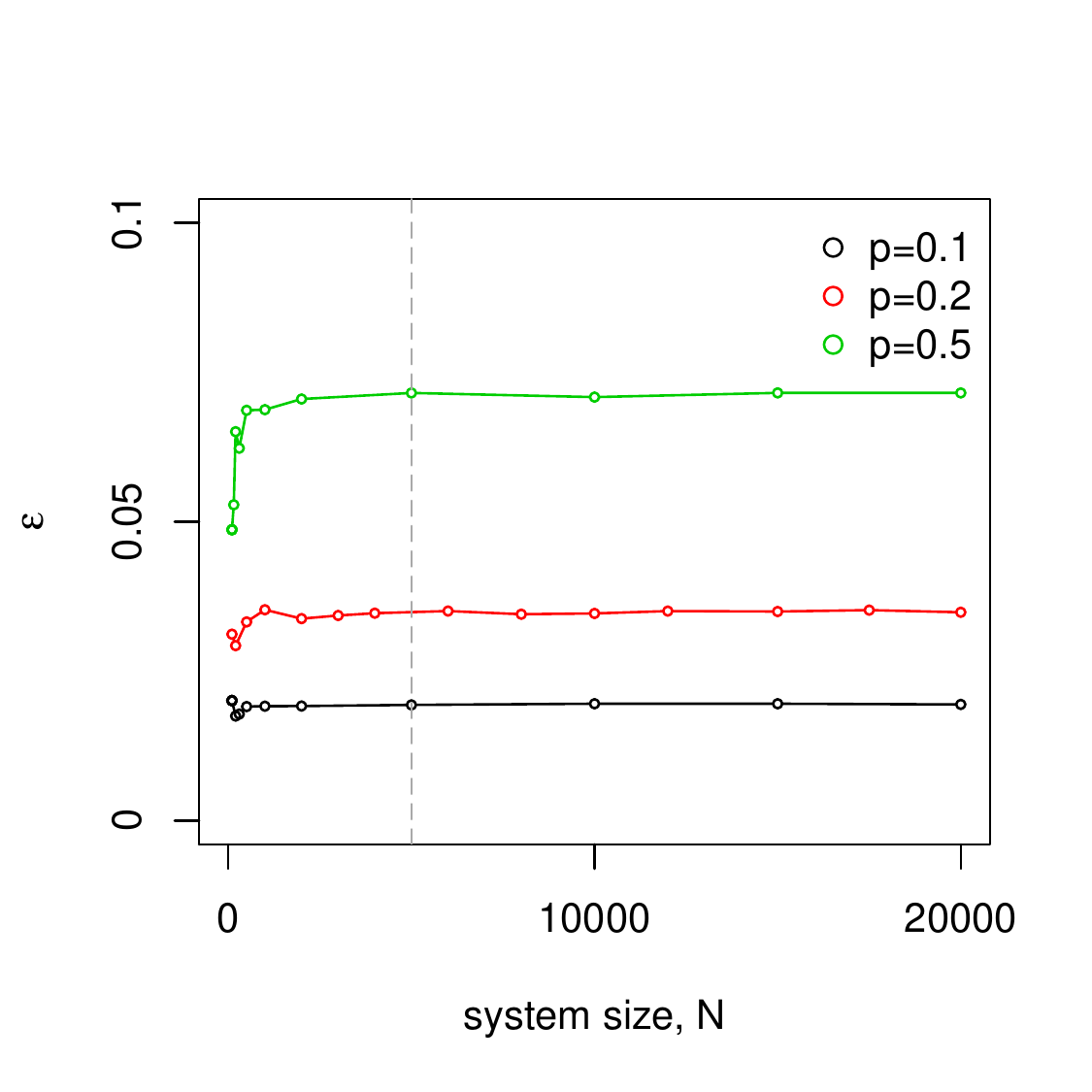}
\caption{ {\bf System size dependence.} 
Simulation for regular graph with $k=3$ (compare Fig.~2b), but varying system size and $p$.
Each curve represents one value of $p$ (marked in legend) and points are simulation results for the transition between $u$-dominant and $w$-dominant states. 
Lines serve as guide to the eye and connect the points obtained by the simulations.
Vertical dashed line indicates $N=5,000$, a value we find sufficient for finite size effects to be considered small.
}
\label{fig:comparison_system_size}
\end{center}
\end{figure}

\noindent
{\bf System size ---} For varying values of the asymmetry parameter $p$, we investigate the dependence of the bifurcation point on system size (Fig.~\ref{fig:comparison_system_size}). 
The results show that for regular graphs already system sizes of approximately $5,000$ nodes are sufficient to reach stable estimates for the critical value of noise.

\begin{figure}[ht!]
\begin{center}
\begin{overpic}[width=5cm,angle=-90,trim= 0cm 0pt 0pt 0pt,clip]{dummy.pdf}
\put(-10,10){\includegraphics[width=8cm,trim= 0cm .7cm 5cm 3.3cm,clip]{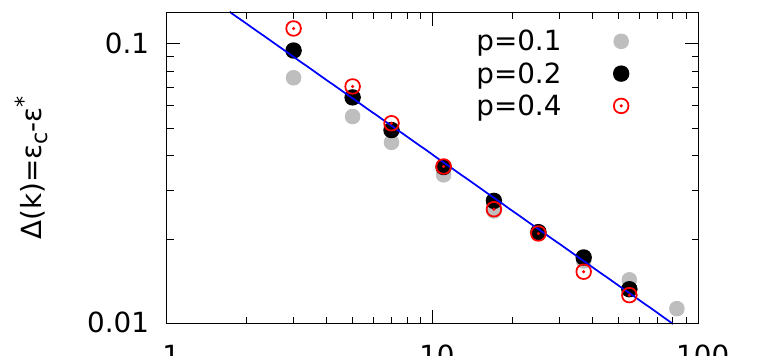}}
\put(-10,-50){\includegraphics[width=8cm,trim= 0cm 0cm 5cm 3.3cm,clip]{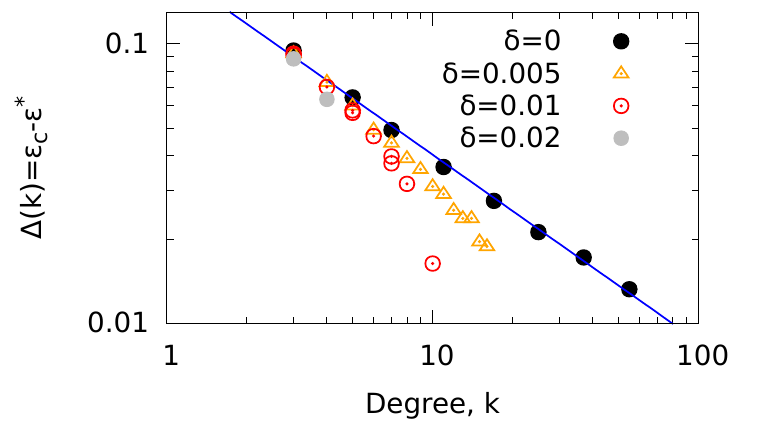}}
\put(-5,55){\bf a}
\put(-5,5){\bf b}
\end{overpic}
\vspace{100pt}
\caption{ {\bf Bifurcation point vs. degree and noise bias.} 
{\bf a}, Difference of fully-connected, infinite system bifurcation point $\varepsilon_c$ and regular graph transition point $\varepsilon^*$ for varying values of degree $k$ and different values of $p$. System sizes: $8,000$ for $p=.1$ and $p=.4$, $4,000$ for $p=.2$.
{\bf b}, System with noise bias, i.e. for $u\rightarrow v$ and $v\rightarrow w$ noisy transitions were increased to $\varepsilon+\delta$ while the opposite direction was decreased to $\varepsilon-\delta$. Values of $\delta$ as indicated in plot.
System sizes: $N=5,000$, except for $\delta=0$, where $N=5,000$ was used.
}
\label{fig:bifurcation_degree}
\end{center}
\end{figure}

\begin{figure}[ht!]
\begin{center}
\includegraphics[width=8cm,trim= 0cm 0cm 0cm 0cm,clip]{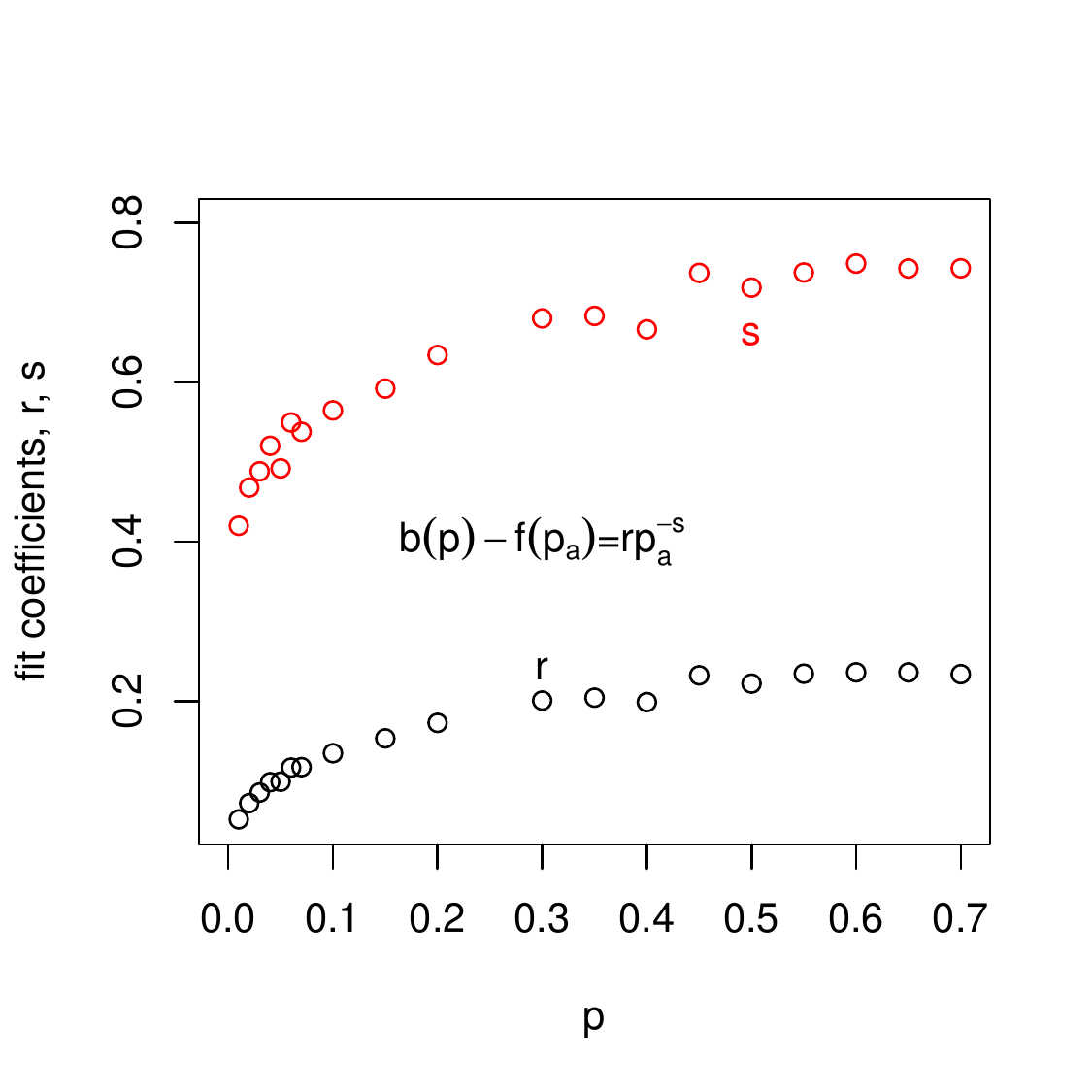}
\caption{ {\bf Fit coefficients for curves shown in Fig.~\ref{fig:bifurcation_degree}a.} 
}
\label{fig:fit_coefficients}
\end{center}
\end{figure}

\begin{figure}[ht!]
\begin{center}
\begin{overpic}[width=5cm,angle=-90,trim= 0cm 0pt 0pt 0pt,clip]{dummy.pdf}
\put(-0,0){\includegraphics[width=6cm,trim= 0cm .55cm 0cm 0cm,clip]{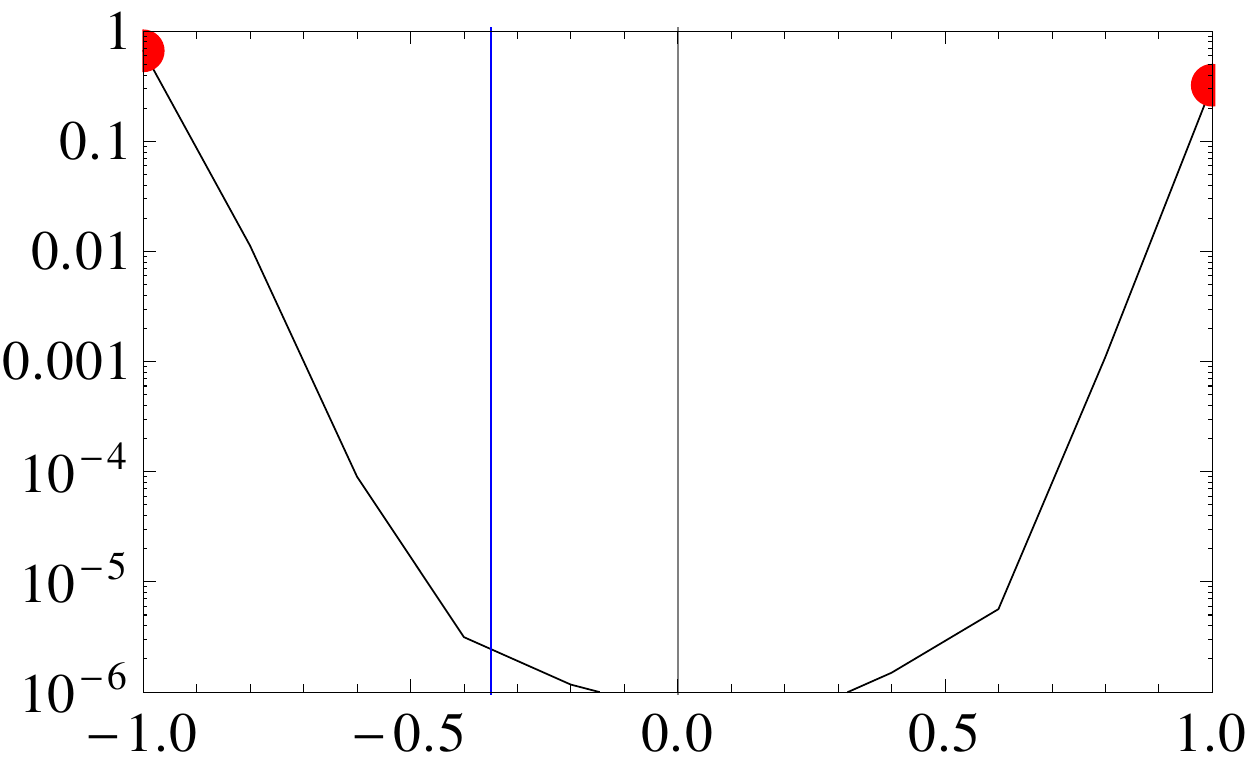}}
\put(-0,-50){\includegraphics[width=6cm,trim= 0cm .55cm 0cm 0cm,clip]{histsSingleStar1.pdf}}
\put(-0,-100){\includegraphics[width=6cm,trim= 0cm .55cm 0cm 0cm,clip]{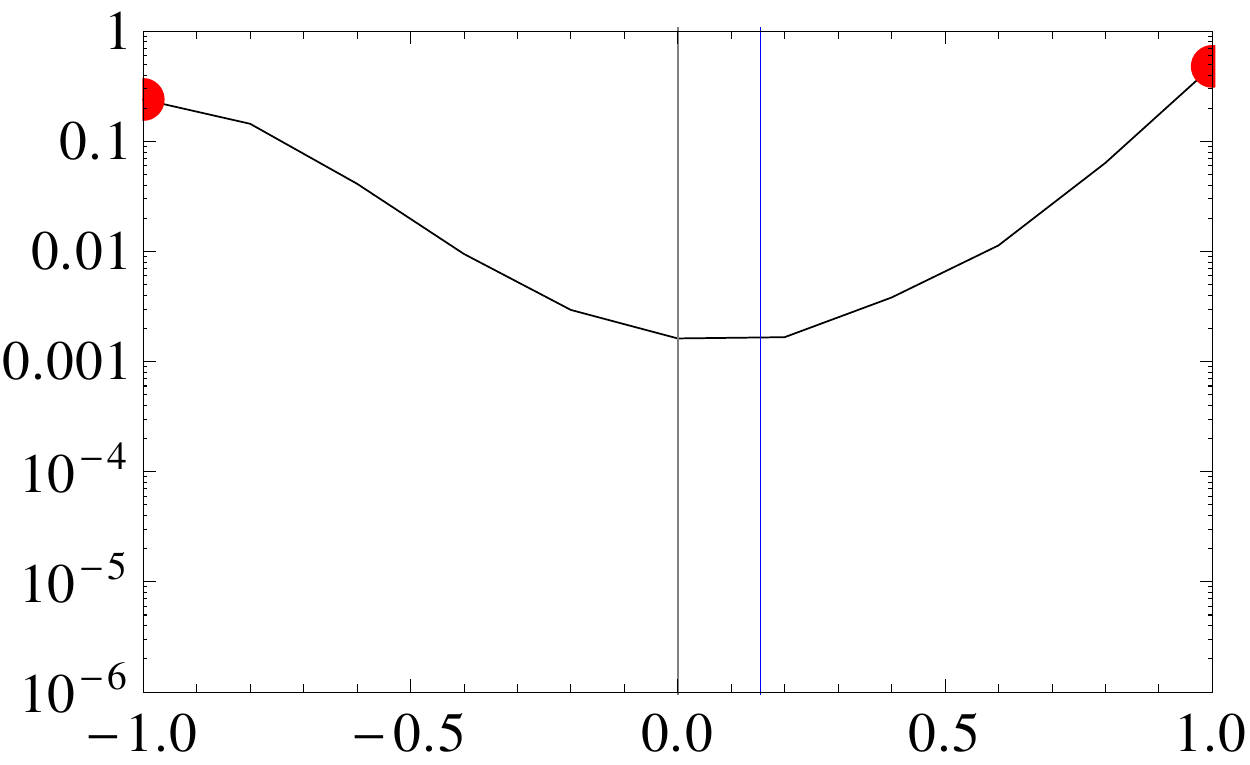}}
\put(-0,-150){\includegraphics[width=6cm,trim= 0cm .55cm 0cm 0cm,clip]{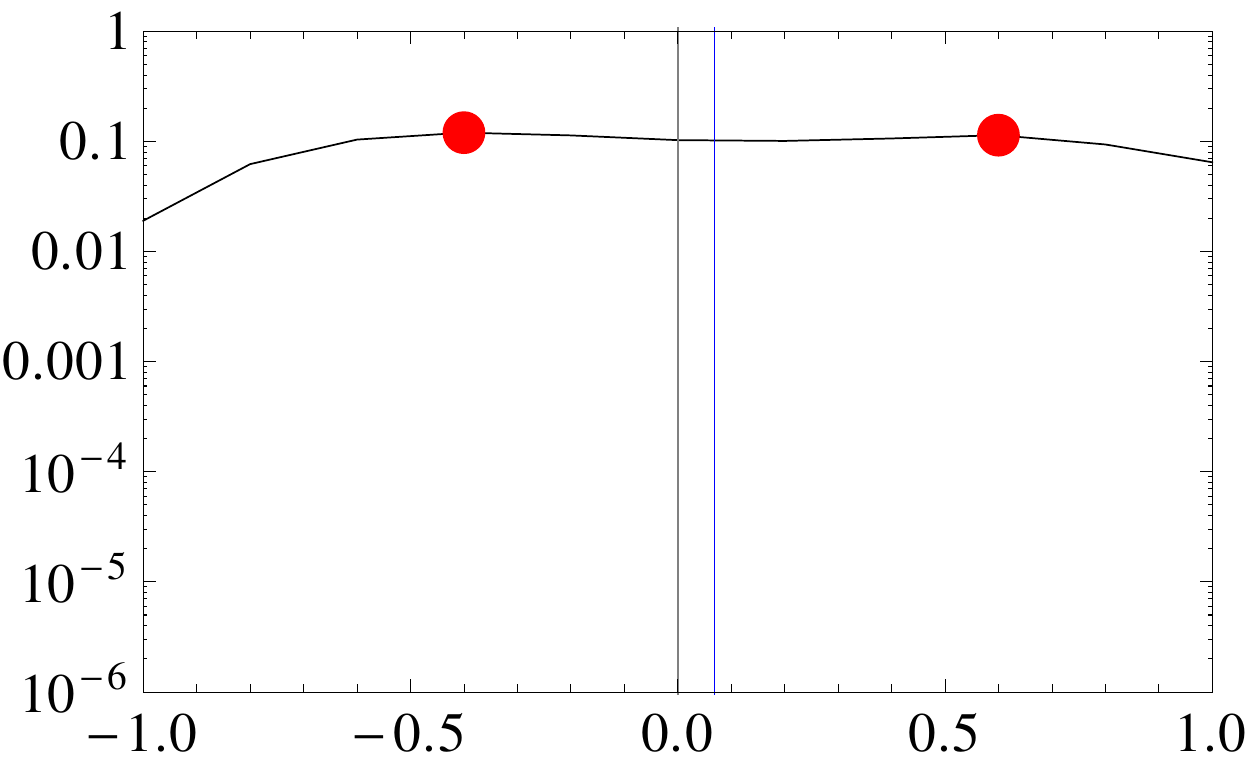}}
\put(-7,-210){\includegraphics[width=6.5cm,trim= 0cm 0cm 0cm 0cm,clip]{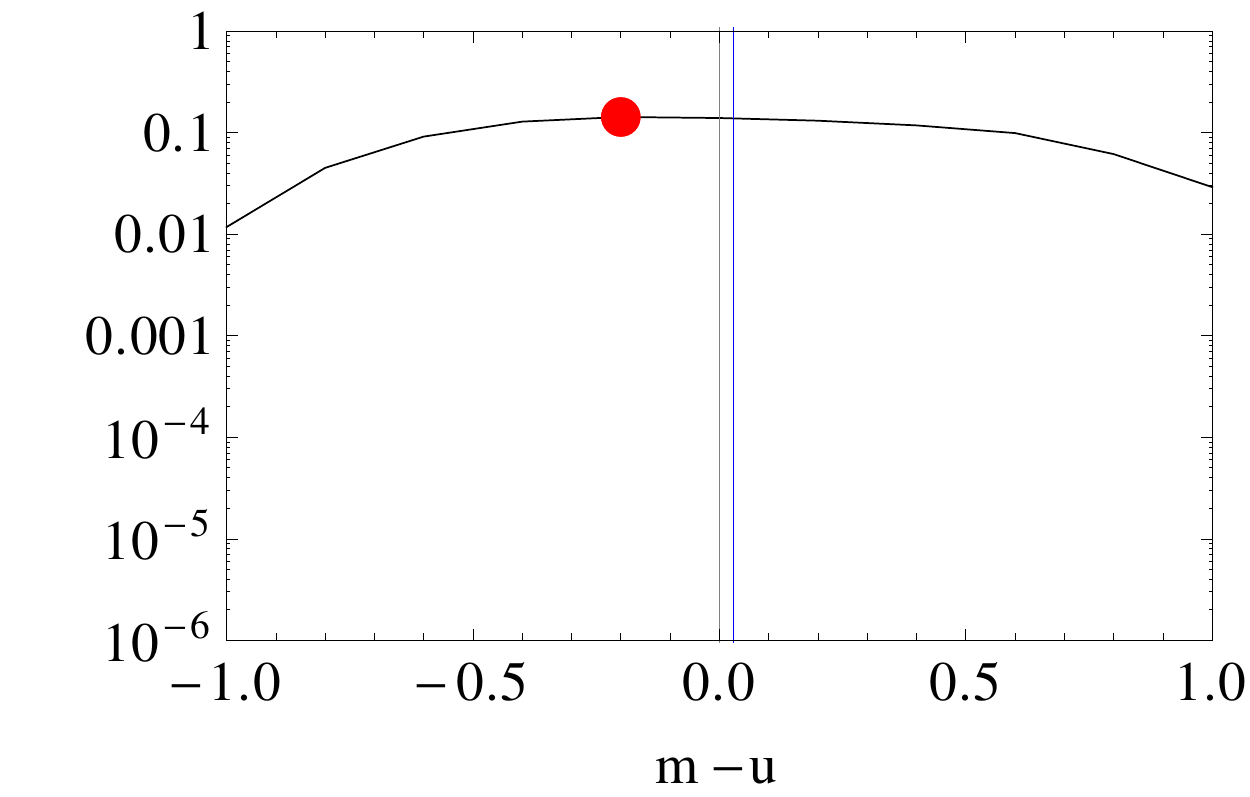}}

\put(-6,43){{\bf a}}
\put(-6,-10){{\bf b}}
\put(-6,-60){{\bf c}}
\put(-6,-110){{\bf d}}
\put(-6,-160){{\bf e}}
\end{overpic}
\vspace{420pt}
\caption{ {\bf Star consisting of five sites.}  
{\bf a---e}, Increasing values of $\varepsilon$.
Examples of probability distribution functions (PDFs) of $w-u$ for all states in a star topology of five sites.
To obtain the PDFs, the value $w-u$ of each star-state was weighted by the corresponding occupation probability of the state.
}
\label{fig:5_star}
\end{center}
\end{figure}

\noindent
{\bf Dependence on connectivity ---}  As pointed out in the main text, even for $p=1$ the bifurcation point is influenced by connectivity (degree). 
As nodes are more and more connected, the bifurcation point approaches the analytical value $\varepsilon_c=\left(1+3/\sqrt{p}\right)^{-1}$ (compare Eq.~\ref{eq:eps_c}).
We define the departure of the transition point $\varepsilon^*$ (dropping explicit reference to $p$ for simplified notation) from the fully-connected bifurcation point $\varepsilon_c$ as
\begin{equation}
 \Delta(k)\equiv\varepsilon_c-\varepsilon^*(k)\;.
 \label{eq:Delta}
\end{equation}
As $\Delta(k)$ will approach zero as $k\rightarrow \infty$, i.e. when approaching the fully-connected system, we can quantify the functional form of the approach.
Fig.~\ref{fig:bifurcation_degree} shows $\Delta(k)$ on a double-logarithmic scale.
For different values of $p$ the curves approach zero as a power-law, with varying coefficients (Fig.~\ref{fig:fit_coefficients}). 

We further explore the effect of biased noise, i.e. we test whether the transition in polarization is still possible when an overall drift towards the $u$-state is present. 
We induce such a drift by modifying the noise rate $\varepsilon$ in such a way that $\varepsilon_-\equiv\varepsilon+\varepsilon'$ is the rate of transition for the reactions $w\rightarrow v$ as well as $v\rightarrow u$, while for the inverse reactions $\varepsilon_+\equiv \varepsilon-\varepsilon'$ (To be clear, note that the symbol $\varepsilon'$ used here should not be confused with the previous symbol in Eqs \ref{eq:branches1} and \ref{eq:branches2}).
The coefficient $\varepsilon'$ thereby describes the drift in the direction of $u$.
For various values of $\varepsilon'$ we show again the dependence on degree $k$ (Fig.~\ref{fig:bifurcation_degree}b).
The plot demonstrates that indeed the transition is present for varying levels of drift. 
However, the curves for $\varepsilon'>0$ are no longer power-law curves, the decay as a function of degree is now faster. 
Even moderate values of connectivity now remove the transition, i.e. connectivity can be seen as supporting the disorder in the system, thereby removing the correlations that previously were necessary for a transition to occur.

\begin{figure}[ht!]
\begin{center}
\includegraphics[width=8cm,trim= 0cm 0cm 0cm 0cm,clip]{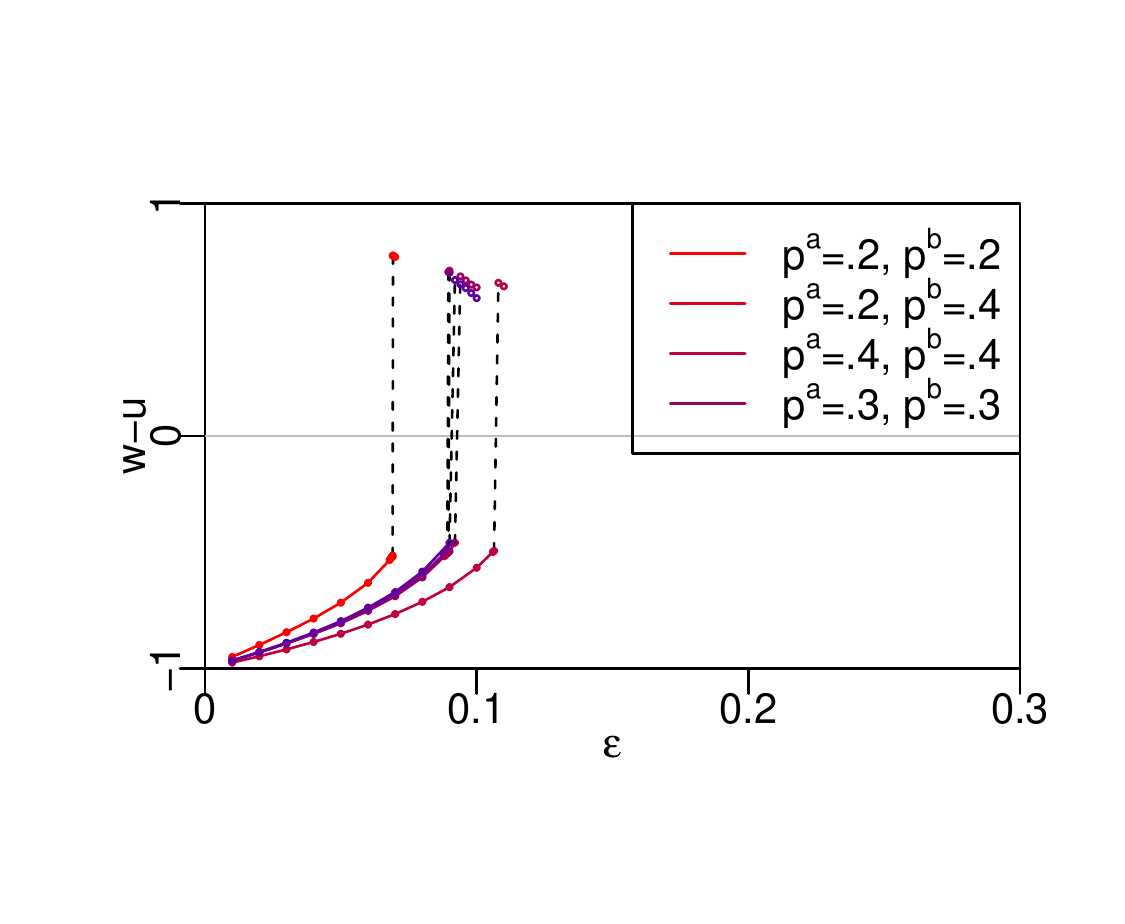}
\caption{{\bf Transition for perfect coupling and various combinations of model parameters.}
Parameters  $p^a$ and $p^b$ correspond to the parameter $p$ in the model (Fig.~1) for the two subsystems $a$ and $b$.
Updating a node in one subsystem will also update the node in the other subsystem.
Curves obtained for subsystems of $N=2,000$ nodes each and degree $k=3$.
Mixing two layers has the main effect of averaging the rates of the two subsystems, seen by comparing the cases of $p^a=.2, p^b=.4$ to $p^a=.3, p^b=.3$. 
} 
\label{fig:mixing_pars}
\end{center}
\end{figure}

Fig.~\ref{fig:mixing_pars} shows the transition for a perfectly coupled two-layer system, i.e. each node exchanges its state between the two layers when it is updated.
We ask, whether different dynamics, defined by differing parameters $p$ in the two layers, can induce differences in the transition of the coupled system. 
The figure shows that the transition for the coupled system can be described well by averaging the parameters within the two individual layers, i.e. a coupled system with parameters $p^a=.2$ (layer a) and $p^b=.4$ (layer b) behaves similarly as a system with both parameters set equal to the average $.3$.

\begin{figure}[ht!]
\begin{center}
\begin{overpic}[width=5cm,angle=-90,trim= 0cm 0pt 0pt 0pt,clip]{dummy.pdf}
\put(-20,-100){\includegraphics[height=10cm,trim= 0cm 0cm 0cm 0cm,clip]{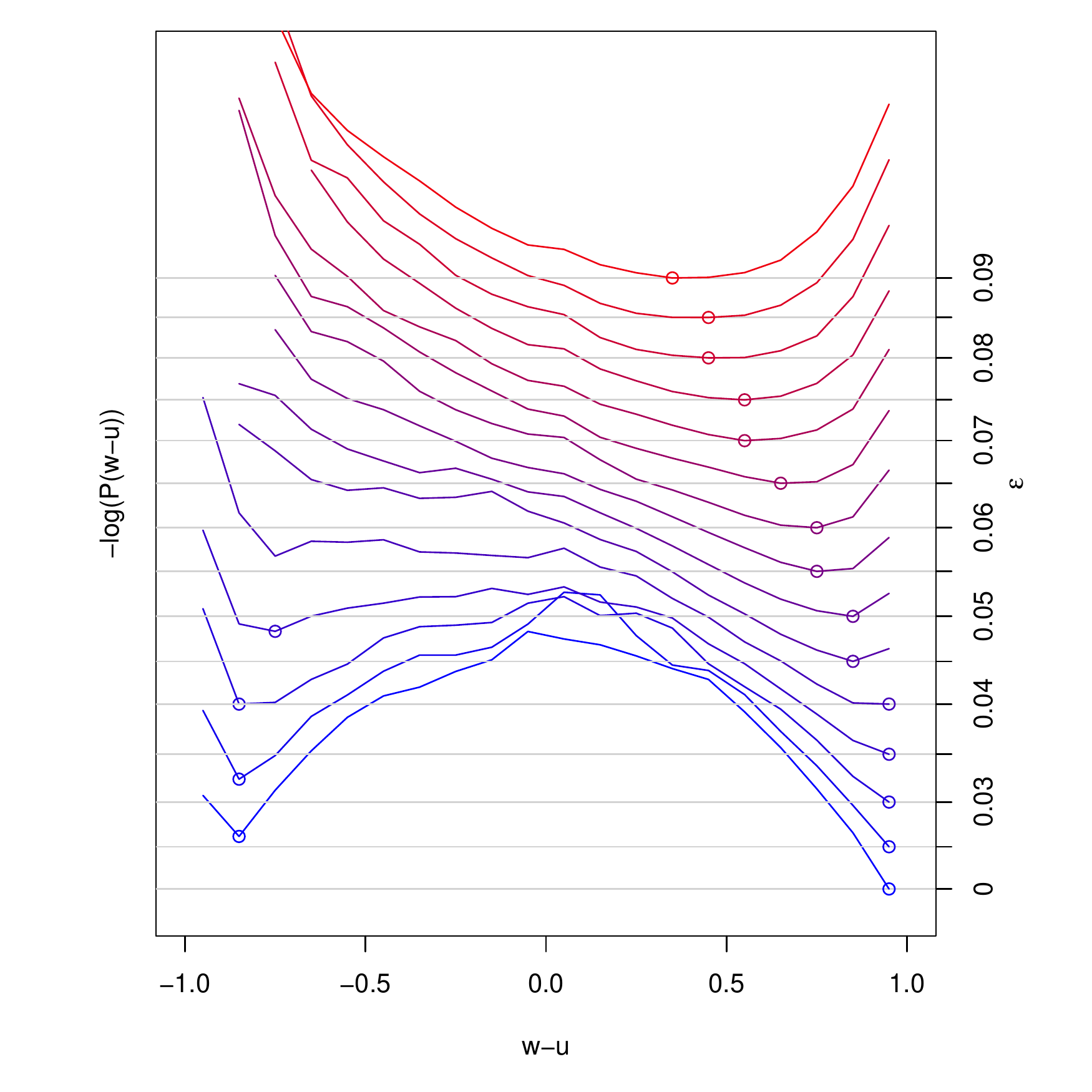}}

\end{overpic}
\vspace{205pt}
\caption{ {\bf Potential wells.}  
For each value of noise $\varepsilon$ long simulations with $n=59$ nodes and $k\approx 3\ll n$ were carried out. 
As these systems are comparably small, sufficiently many fluctuations between the quasi-stable states at negative and positive $M$ do occur to compute the probability density function $P(M=w-u)$. 
Curves ranging from blue to red show the function $-log(P(w-u))$ for a range of $\varepsilon$. 
Inspecting these functions shows that the minimum at $M<0$ becomes increasingly weak and finally disappears. The function obtains an overall slant towards positive $M$.
}
\label{fig:ln_p}
\end{center}
\end{figure}

{\bf Verifying abruptness of transition ---} To prove that the sparsely-connected systems, e.g. shown in Fig.~2, in fact perform an abrupt transition, we map out the occupation probability density of polarization $P(w-u)$ for each value of noise. 
To obtain the occupation probability density, we have performed long simulations, requiring that the system transition sufficiently many times between the two extremal states within the bistable regime, i.e. the entire range of states must be sufficiently explored.
As transitions become exceedingly rare for larger systems, we were only able to compute the spectrum for systems up to $59$ sites. 
Larger systems gave very noisy results, but slightly smaller ones were qualitatively consistent with the results we show here.
The function $-log(P(w-u))$ can be seen as describing the ``potential function'' of the correlated system (Fig.~\ref{fig:ln_p}). 
Notably, a bifurcation is defined here as the transition from a bimodal to a unimodal regime, associated with the disappearance of one stable fixed point as noise is increased.
Indeed, the figure shows the presence of two potential minima, corresponding to stable fixed points, for low noise but only one remaining stable fixed point ($w-u>0$) for larger noise. 
The disappearance of the fixed point $w-u<0$ is abrupt in the sense that the two fixed points do not merge as noise is increased. 
Rather, the potential minimum at low $w-u$ gradually becomes more and more shallow and finally disappears altogether.

{\bf Variants of coupled two-layer systems ---} 
We have further investigated several ways of coupling two-layer systems. 
Beyond the case shown in the main text (Fig.~4), we here show additional cases: 
\begin{itemize}
 \item Not a fraction of nodes is permanently coupled, but all nodes are coupled at a reduced rate $c_r$. The rate is a probability that states how likely it is that one node will couple its state to the other layer after it has been updated (Fig.~\ref{fig:coupling_2layer_3panel}b).
 \item The product of the two coupling coefficients $c_rc_f$ is held constant, and different values of $c_r$ are explored (Fig.~\ref{fig:coupling_2layer_3panel}c).
 \item Another type of game is explored, where the two layers have similar hierarchy of parameters ($p<1$ in both layers), but the magnitude of this parameter varies between the two layers (Fig.~\ref{fig:coupling_2layer_3panel}). 
\end{itemize}

\begin{figure}[ht!]
\begin{center}
\begin{overpic}[width=5cm,angle=-90,trim= 0cm 0pt 0pt 0pt,clip]{dummy.pdf}

\put(-15,11){\includegraphics[width=9.5cm,trim= 0cm 1.0cm 0cm 0cm,clip]{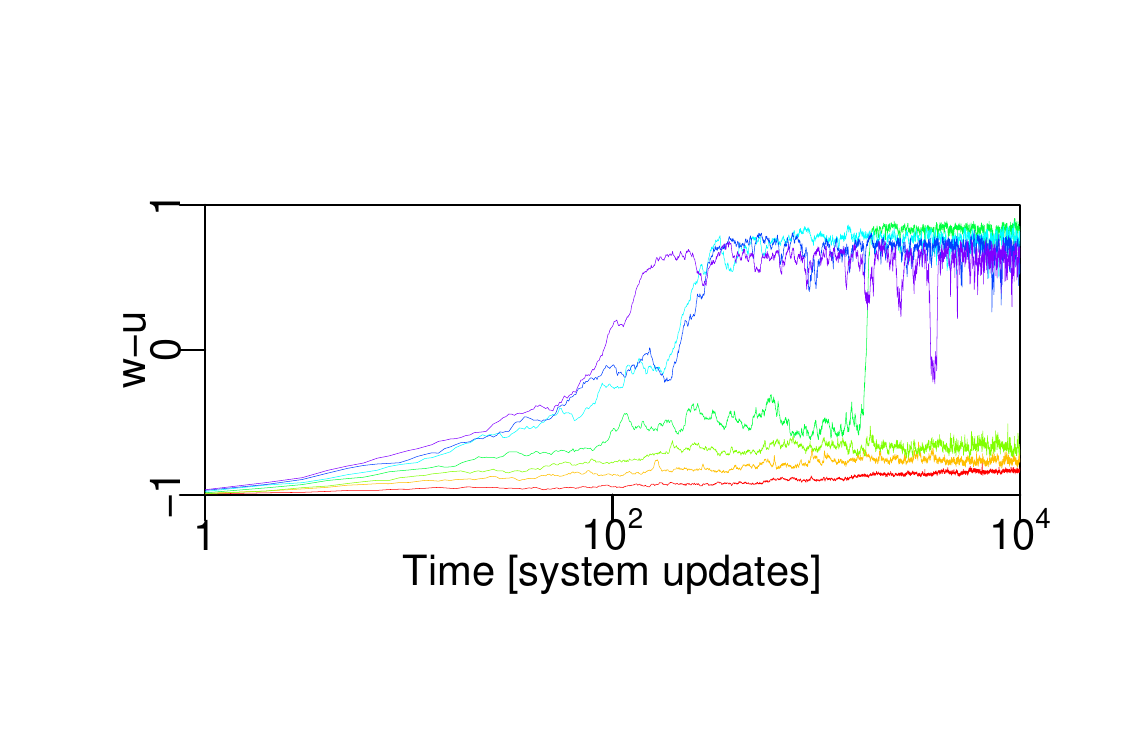}}
\put(-15,-20){\includegraphics[width=9.5cm,trim= 0cm 2.5cm 0cm 0cm,clip]{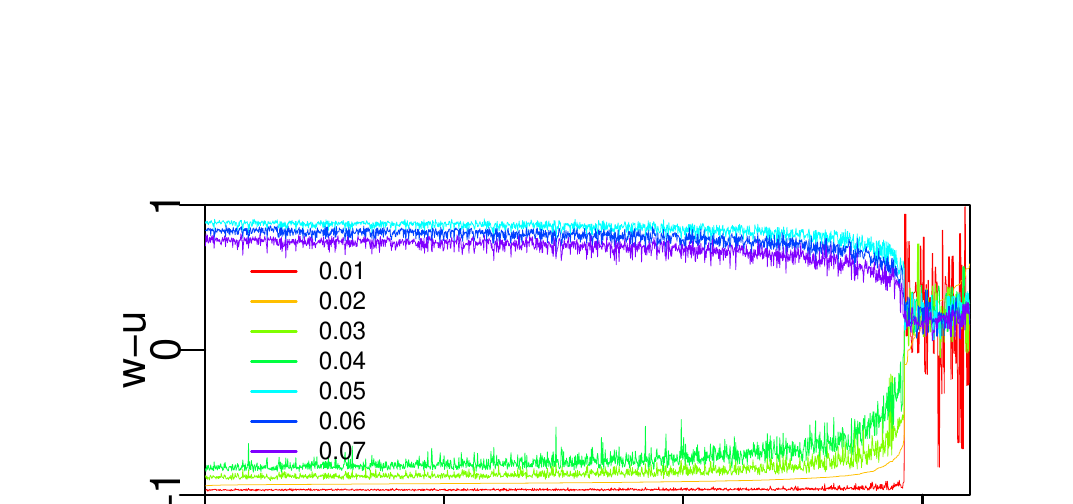}}
\put(-15, -55){\includegraphics[width=9.5cm,trim= 0cm 1.5cm 0cm 0cm,clip]{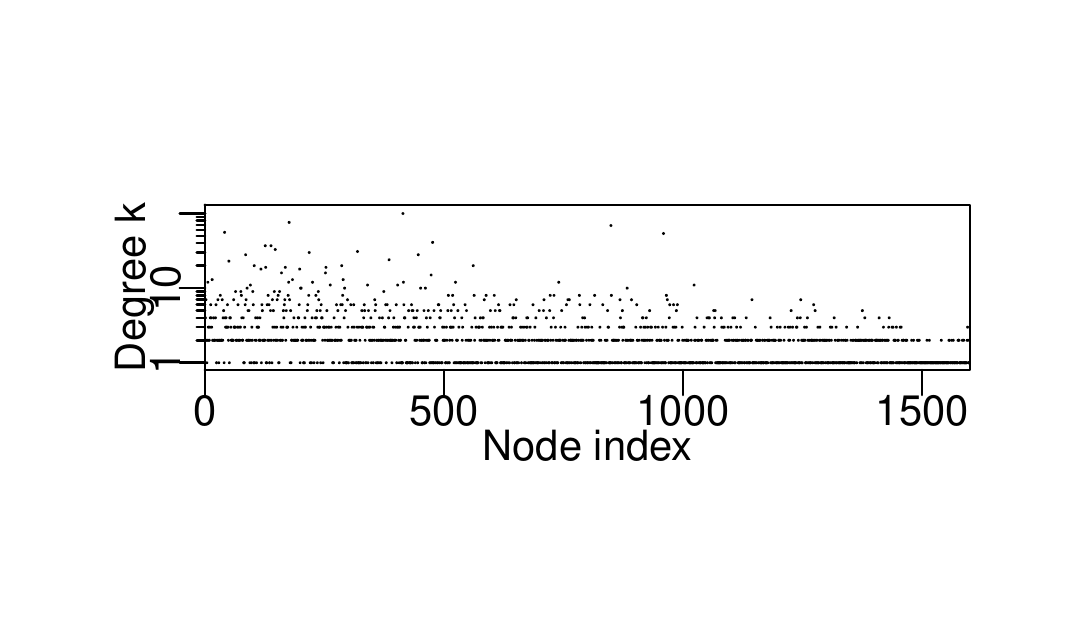}}


\put(13,59){{\bf a}}
\put(13,8){{\bf b}}
\put(13,-25){{\bf c}}
\end{overpic}
\vspace{100pt}
\caption{ {\bf Networks and node states for a synthetic scale free network.}  
Analogous to Fig.~3 but for a synthetic scale free network consisting of 1,600 nodes and average degree $\langle k\rangle\approx 3$.
Note that the timeseries, previously shown in the inset, is now shown as a separate panel in (a).
}
 \label{fig:networks_synthetic}
\end{center}
\end{figure}

\begin{figure}[ht!]
\begin{center}
\begin{overpic}[width=5cm,angle=-90,trim= 0cm 0pt 0pt 0pt,clip]{dummy.pdf}
\put(0,   0){\includegraphics[width=7cm,trim= 0cm 2.5cm 0cm 0cm,clip]{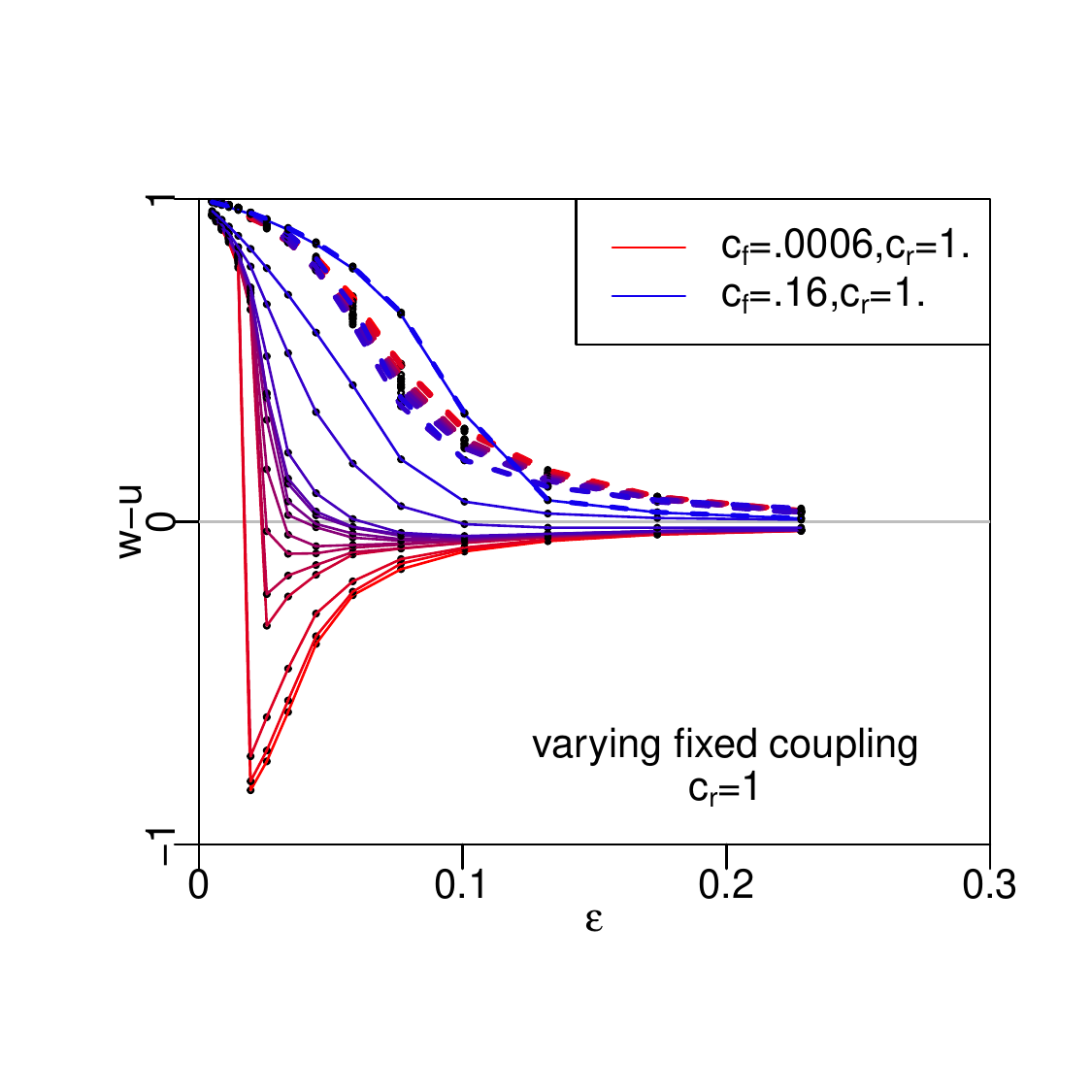}}
\put(0, -60){\includegraphics[width=7cm,trim= 0cm 2.5cm 0cm 0cm,clip]{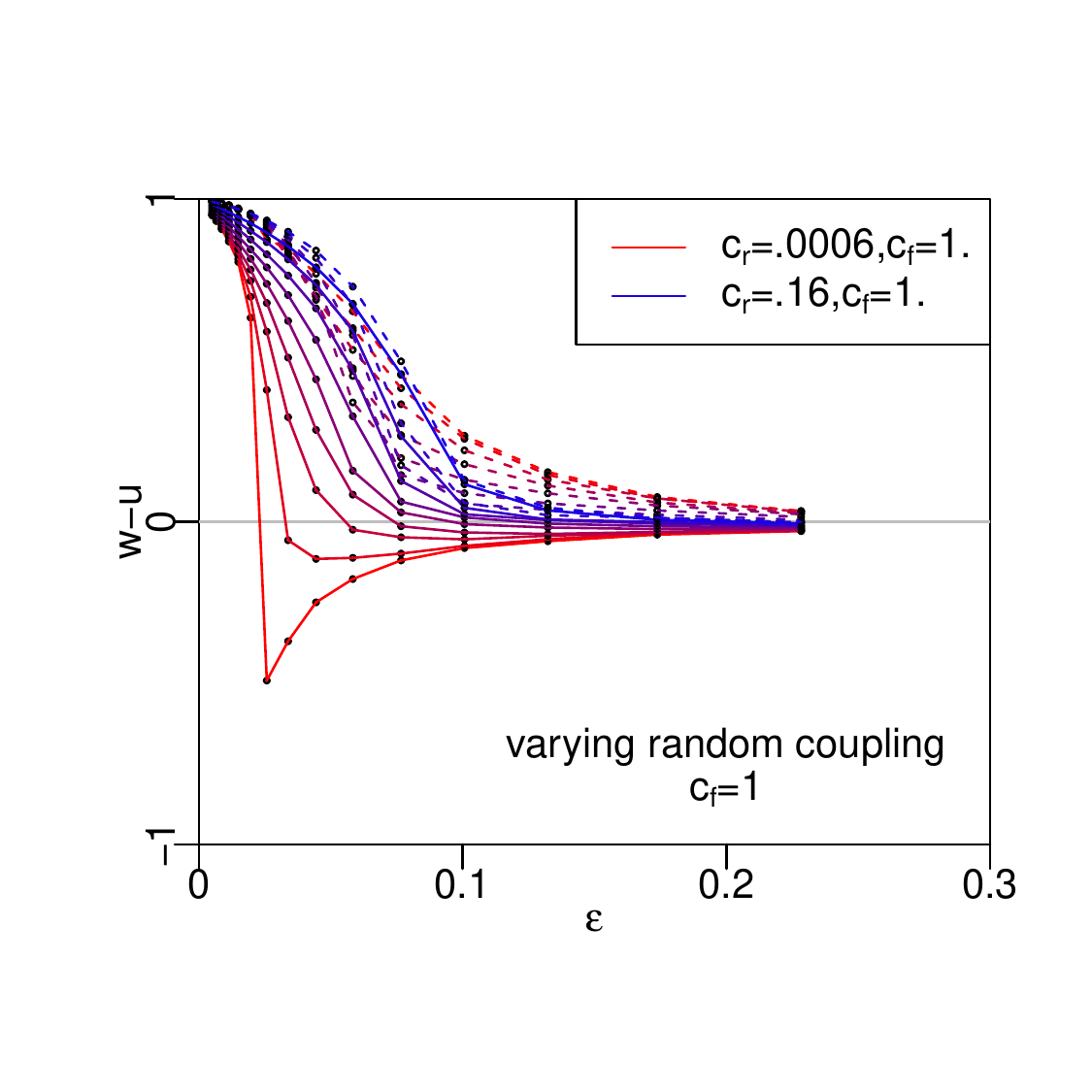}}
\put(0,-120){\includegraphics[width=7cm,trim= 0cm 2.5cm 0cm 0cm,clip]{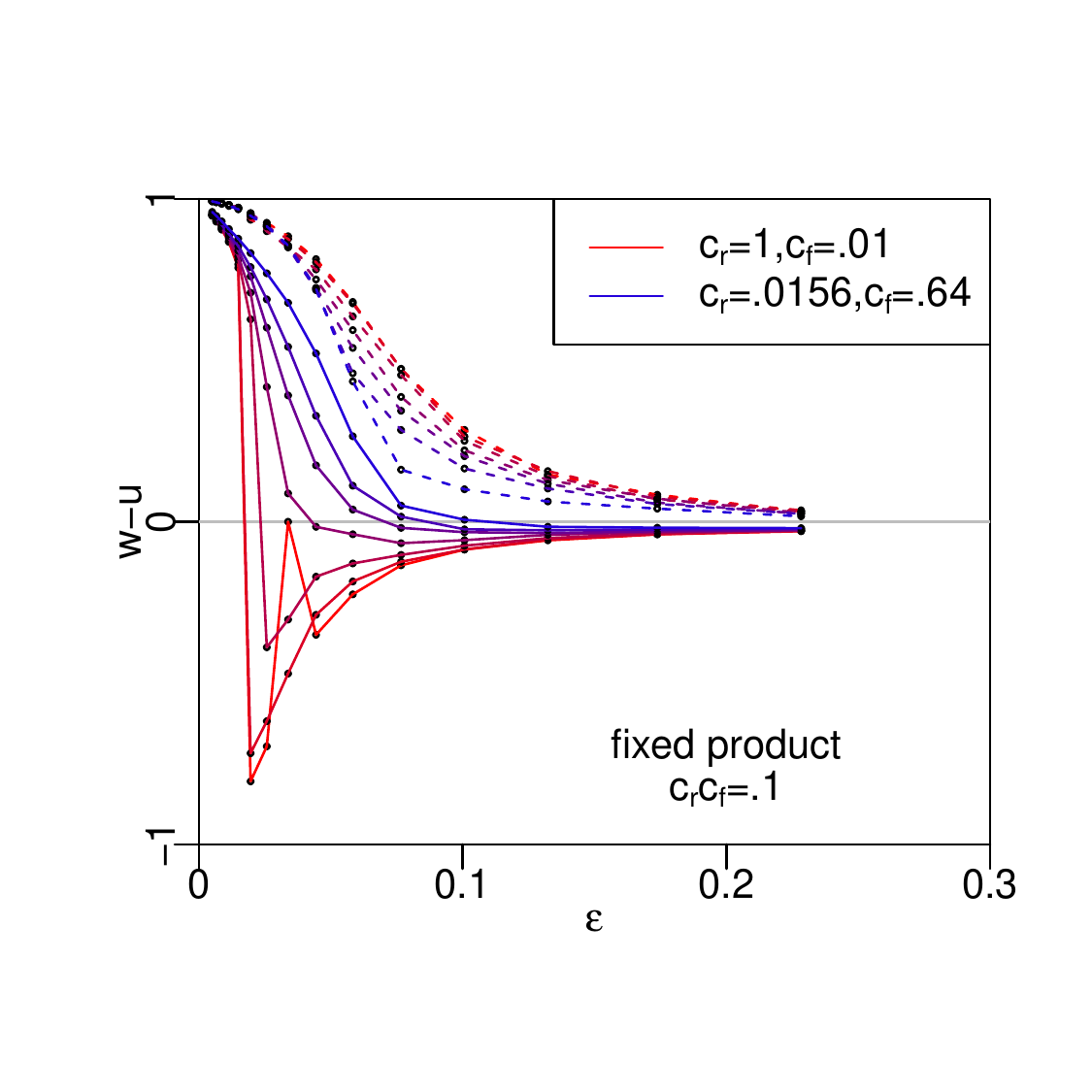}}
\put(0,-202){\includegraphics[width=7cm,trim= 0cm 0cm 0cm 0cm,clip]{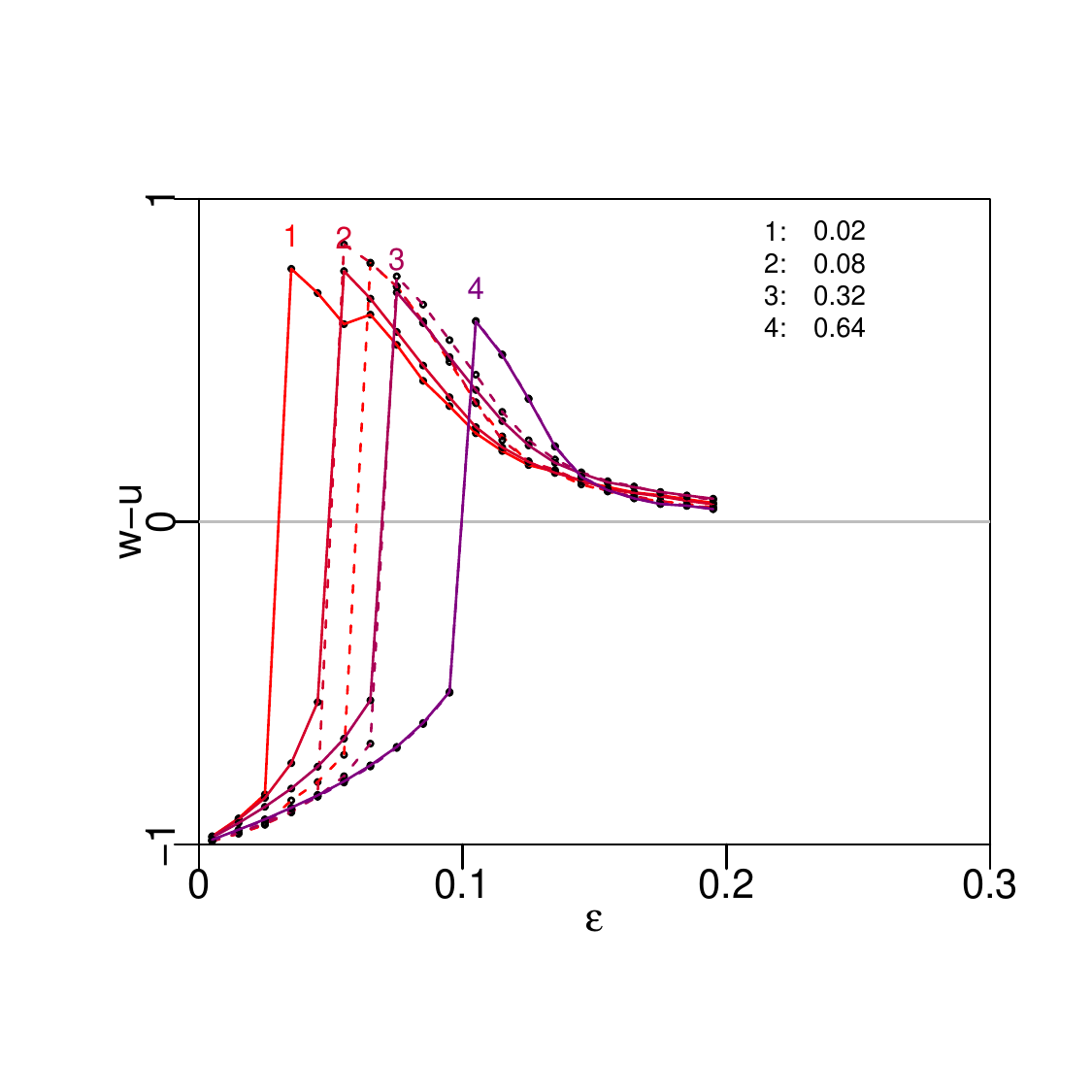}}
\put(36,-188){\includegraphics[width=4cm,trim= 0cm 0cm 0cm 0cm,clip]{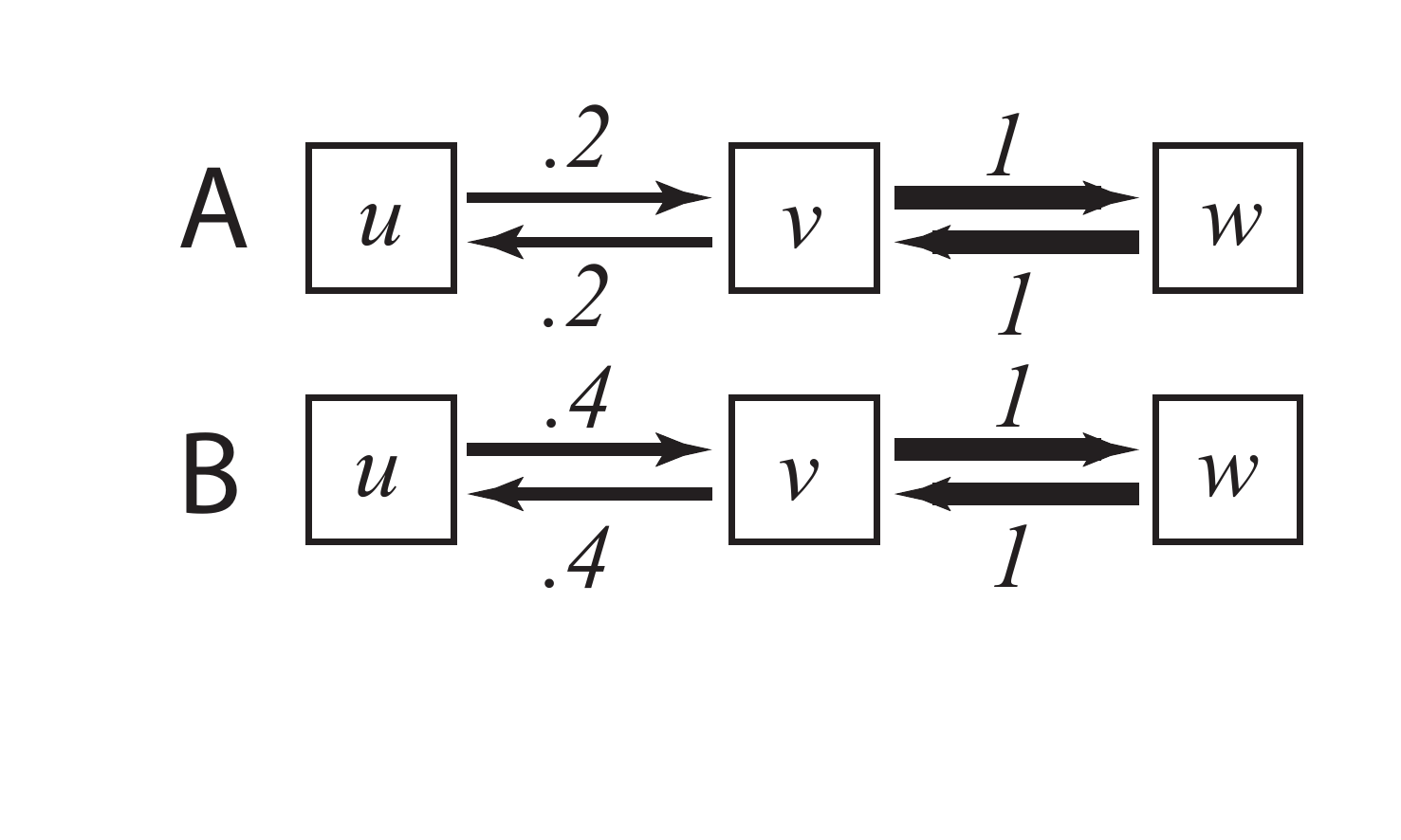}}
\put(0,55){{\bf a}}
\put(0,-5){{\bf b}}
\put(0,-65){{\bf c}}
\put(0,-125){{\bf d}}
\end{overpic}
\vspace{385pt}
\caption{ {\bf Effect of coupling a two-layer system.}  
{\bf a}, A fraction $c_f$ is coupled between two layers of a regular ($k=3$) network. 
There is no randomness of coupling noise ($c_r=1$), i.e. at each update of a given coupled node, the state of this node in one layer is copied to that in the other.
The parameter $c_f$ is varied as shown in the legend.
{\bf b}, Similar to (a) but now keeping $c_f=1$ maximal, i.e. all nodes are capable of copying their states. Now, $c_r$ is varied as shown in the legend, i.e. the probability of copying the state in one layer to the other layer.
{\bf c}, Similar to (a) but keeping the product $c_rc_f=.1$ fixed.
Parameters of different curves as shown in legend.
{\bf d}, Coupling two systems as shown in panel inset.
Solid (dashed) lines represent systems A and B, respectively. 
$c_r=1$. 
Numbers near lines and in legend represent the fraction of nodes permanently coupled ($c_f$).
System sizes in all panels: $N=2,000$.
}
\label{fig:coupling_2layer_3panel}
\end{center}
\end{figure}

\bibliography{references_parrondo}
\bibliographystyle{apsrev4-1}